\documentclass[apj]{emulateapj}

\bibpunct{(}{)}{;}{a}{}{,}

\slugcomment{Accepted to the ApJ}

\shorttitle{ {\it HST} PL DOG Morphologies}
\shortauthors{Bussmann et al.}

\begin{document}

\title{$HST$ Morphologies of $z \sim 2$ Dust Obscured Galaxies I: Power-law
Sources}

\author{R. S. Bussmann\altaffilmark{1}, Arjun Dey\altaffilmark{2}, J.
Lotz\altaffilmark{2,3}, L.
Armus\altaffilmark{4}, K. Brand\altaffilmark{5}, M. J. I.
Brown\altaffilmark{6}, V.  Desai\altaffilmark{4}, P.
Eisenhardt\altaffilmark{7}, J. Higdon\altaffilmark{8}, S.
Higdon\altaffilmark{8}, B. T.  Jannuzi\altaffilmark{2}, E. Le
Floc'h\altaffilmark{9}, J.  Melbourne\altaffilmark{10}, B. T.
Soifer\altaffilmark{4,10}, D. Weedman\altaffilmark{11}}

\altaffiltext{1}{Steward Observatory, Department of Astronomy, University of
Arizona, 933 N. Cherry Ave., Tucson, AZ 85721; rsbussmann@as.arizona.edu}
\altaffiltext{2}{National Optical Astronomy Observatory, 950 N. Cherry Ave., Tucson, AZ 85719}
\altaffiltext{3}{NOAO Leo Goldberg Fellow}
\altaffiltext{4}{Spitzer Science Center, California Institude of Technology, MS
220-6, Pasadena, CA 91125}
\altaffiltext{5}{Giacconi Fellow, Space Telescope Science Institute, Baltimore, MD 21218}
\altaffiltext{6}{School of Physics, Monash University, Clayton, Victoria 3800,
Australia}
\altaffiltext{7}{Jet Propulsion Laboratory, California Institute of Technology,
MC 169-327, 4800 Oak Grove Drive, Pasadena, CA 91109}
\altaffiltext{8}{Georgia Southern University, P.O. Box 8031, Statesboro, GA}
\altaffiltext{9}{Spitzer Fellow, Institute for Astronomy, University of Hawaii, Honolulu, HI 96822}
\altaffiltext{10}{Caltech Optical Observatories, California Institute of
Technology, Pasadena, CA 91125}
\altaffiltext{11}{Astronomy Department, Cornell University, Ithaca, NY 14853}


\begin{abstract}

We present high spatial resolution optical and near-infrared imaging obtained using the ACS, WFPC2 and NICMOS cameras aboard the {\it Hubble Space Telescope} of 31 24$\mu$m--bright $z\approx2$ Dust Obscured Galaxies (DOGs) identified in the Bo\"otes Field of the NOAO Deep Wide-Field Survey. Although this subset of DOGs have mid-IR spectral energy distributions dominated by a power-law component suggestive of an AGN, all but one of the galaxies are spatially extended and not dominated by an unresolved component at rest-frame UV or optical wavelengths. The observed $V-H$ and $I-H$ colors of the extended components are $0.2-3$~magnitudes redder than normal star-forming galaxies.  All but 1 have axial ratios $>0.3$, making it unlikely that DOGs are composed of an edge-on star-forming disk. We model the spatially extended component of the surface brightness distributions of the DOGs with a S\'ersic profile and find effective radii of $1-6$~kpc.  This sample of DOGs is smaller than most sub-millimeter galaxies (SMGs), but larger than quiescent high-redshift galaxies. Non-parametric measures (Gini and M$_{20}$) of DOG morphologies suggest that these galaxies are more dynamically relaxed than local ULIRGs. We estimate lower limits to the stellar masses of DOGs based on the rest-frame optical photometry and find that these range from $\sim10^{9-11}~ M_\sun$. If major mergers are the progenitors of DOGs, then these observations suggest that DOGs may represent a post-merger evolutionary stage.


\end{abstract}



\section{Introduction} \label{sec:intro} 

One of the most important questions concerning the evolution of galaxies is
when and how the most massive galaxies formed.  It has been known since the
analysis of the {\it InfraRed Astronomical Satellite} ($IRAS$) data that in the
local universe the most bolometrically luminous galaxies have their spectral
energy distributions (SEDs) dominated by infrared (IR) light
\citep{1986ApJ...303L..41S}, suggesting that these systems are highly obscured
by dust, which absorbs ultra-violet (UV) and optical light and re-radiates it
in the IR.  While these ultra-luminous IR galaxies (ULIRGs) are rare in the
local universe, they become an increasingly important phenomenon at high
redshift
\citep[e.g.,][]{2001A&A...378....1F,2005ApJ...632..169L,2005ApJ...630...82P}.

Following the launch of the {\it Spitzer Space Telescope}, numerous
investigators have identified and studied populations of high redshift galaxies
that are IR-bright yet optically faint
\citep{2004ApJS..154...60Y,2005ApJ...622L.105H,2006ApJ...651..101W,2008ApJ...672...94F,2008ApJ...677..943D}.  
In particular, \citet{2008ApJ...677..943D} and \citet{2008ApJ...672...94F}
present a simple and economical method for selecting these systems using only
$R$-band and 24$\mu$m Multiband Imaging Photometer for Spitzer
\citep[MIPS;][]{2004ApJS..154...25R} data.  \citet{2008ApJ...677..943D} employ
a color cut of $R - [24] > 14$ (Vega magnitudes; $\approx$$\, F_\nu (24\mu {\rm
m})/F_\nu (R) > 1000$) to identify objects they call Dust Obscured Galaxies
(DOGs) in the Bo\"{o}tes field of the NOAO Deep Wide-Field Survey
(NDWFS\footnote{http://www.noao.edu/noaodeep}; Jannuzi et al., in prep.; Dey et
al., in prep.).

The broadband photometry, redshift distribution, and number density of the DOGs
imply that they are undergoing an extremely luminous, short-lived phase of
stellar bulge and nuclear black hole growth and may be the progenitors of the
most luminous ($\sim$4$L^*$) present-day galaxies.  Ground-based photometry
from the NDWFS suggests magnitudes of $R\approx24-27$, $I\approx24-26$, and
$K\approx17.5-20.5$ for the sample of DOGs with $F_\nu (24\mu m) > 0.3 \:$mJy.
DOGs are relatively rare, with a surface density of
$\approx$0.089~arcmin$^{-2}$ for sources with $F_{24} > 0.3 \:$mJy.
Spectroscopic redshifts determined for a sub-sample of DOGs using the Deep
Imaging Multi-Object Spectrograph \citep[DEIMOS;][]{2003SPIE.4841.1657F} and
the Low Resolution Imaging Spectrometer \citep[LRIS;][]{1995PASP..107..375O} on
the telescopes of the W.~M.~Keck Observatory (43 DOGs), as well as the Infrared
Spectrometer \citep[IRS;][]{2004ApJS..154...18H} on {\it Spitzer} (43 DOGs)
have shown that the DOGs have a redshift distribution centered on $z \approx 2$
with a dispersion of $\sigma_z \approx 0.5$.  While DOGs are rare, they are
sufficiently luminous that they contribute up to one-quarter of the total IR
luminosity density from all $z \sim 2$ galaxies, and constitute the bulk of
ULIRGs at $z \sim 2$ \citep{2008ApJ...677..943D}.  

Based on their observed properties, \citet{2008ApJ...677..943D} suggest DOGs
may represent a transition stage between sub-millimeter-selected galaxies
(SMGs) and un-obscured quasars or galaxies.  Evidence in support of this
scenario is that DOGs and SMGs have similar space densities and clustering
properties \citep{2008arXiv0810.0528B}.  An important test of this scenario is
to study their morphologies with high spatial resolution imaging.  For example,
one of the primary motivations for the merger-driven scenario for the formation
of ULIRGs is their disturbed structure at optical wavelengths
\citep{1988ApJ...325...74S}.  Studies of numerical simulations of galaxy
mergers have suggested that they can produce very red, luminous systems that
are highly dust-obscured \citep{2006ApJ...637..255J}.  Recently,
\citet{2008arXiv0805.1246L} have applied non-parametric methods of quantifying
galaxy morphologies to similar merger simulations and have found that mergers
are most easily identified during the first pass and at the final coalescence
of their nuclei.

In addition to identifying merger activity, morphological information can
constrain the size-scale of the emitting region.  Sources with active star
formation on several kiloparsec (kpc) scales have larger sizes than objects
dominated by an Active Galactic Nucleus (AGN) or a very compact, nuclear
starburst.  Studies of Distant Red Galaxies (DRGs) have shown a relation
between star formation and size at rest-frame optical wavelengths, in the sense
that quiescent DRGs are all very compact with effective radii ($R_{\rm eff}$)
less than 1~kpc, while active DRGs tend to be more extended \citep[$1 < R_{\rm
eff} < 10 \:$kpc;][]{2007ApJ...656...66Z,2007ApJ...671..285T}.  Analysis of
SMGs in GOODS-N shows extended emission on scales of 5-15~kpc
\citep{2005MNRAS.358..149P}.  Recent NICMOS imaging of a sample of 33 high-$z$
ULIRGs by \citet{2008arXiv0802.1050D} has shown these extreme objects (which
are similar in their selection criteria to DOGs) to have effective radii in the
range $\sim$1.5-5~kpc.  About half of their sample shows signs of interactions,
but only 2 are merging binaries with a luminosity ratio $\leq$3:1, i.e.,
qualifying as major mergers.  

High spatial resolution imaging of the DOGs is essential to understanding their
relation to other galaxy populations as well as their role in galaxy evolution
in general.  We have begun an effort to obtain high resolution imaging using
laser guide star and natural guide star adaptive optics on the Keck telescopes.
These (on-going) efforts have resulted in high resolution $K$-band images of a
handful of DOGs found near bright stars \citep[][, in
prep.]{2008AJ....136.1110M}.  A complementary method of obtaining deep, high
spatial resolution imaging is with the {\it Hubble Space Telescope} ({\it
HST}).  With the Advanced Camera for Surveys (ACS) and the Wide Field Planetary
Camera~2 (WFPC2), we can probe the rest-frame UV emission of the DOGs that is
sensitive to the ionizing sources associated with on-going star formation.
Meanwhile, NICMOS data allow the study of the rest-frame optical morphology,
which better traces the stellar mass and dust-enshrouded AGN.

In this paper we present ACS/WFPC2 and NICMOS images of 31 DOGs and analyze
their morphologies.  The DOGs studied in this paper have spectroscopic
redshifts from either {\it Spitzer}/IRS 
DEIMOS/LRIS, were selected primarily based on their large 24$\mu$m flux
densities (F$_{24\mu m} > 0.8 \:$mJy), and have power-law SEDs in the mid-IR.
In a future paper, we will study a sample of DOGs with fainter 24$\mu$m flux
densities that have mid-IR bump SEDs (Bussmann et al., in prep.).  In
section~\ref{sec:data} we detail the sample selection, observations, and data
reduction.  Section~\ref{sec:morphmeth} contains a description of the methods
we use in our morphological analysis, and in section~\ref{sec:results}, we
report the results this analysis.  In section~\ref{sec:disc}, we estimate some
intrinsic properties of the DOGs in our sample and we compare our findings with
what is seen in other high redshift galaxy populations.  Finally, we present
our conclusions in section~\ref{sec:conclusions}.

Throughout this paper we assume $H_0=$70~km~s$^{-1}$~Mpc$^{-1}$, $\Omega_{\rm
m} = 0.3$, and $\Omega_\lambda = 0.7$.  At $z=2$, this results in
8.37~kpc/$\arcsec$.

\section{Data}\label{sec:data}

In this section, we describe our sample selection and give details regarding
the {\it HST} observations and our data reduction procedure, as well how we
measure our photometry.  Finally, we show postage stamp images and provide a
brief qualitative description of each target.

\subsection{Sample Selection} \label{sec:sample}

As outlined in section~\ref{sec:intro}, a sample of $\approx$2600 DOGs from
\citet{2008ApJ...677..943D} was originally identified using the 9.3 deg$^2$
Bo\"{o}tes Field of the NDWFS.  For details of the selection criteria and
photometric analysis, we refer the reader to \citet{2008ApJ...677..943D}.  In
this paper, we analyze {\it HST} imaging from program HST-GO10890 of 31 of the
brightest DOGs at 24$\mu m$ (all have $F_{24\mu {\rm m}} > 0.8 \:$mJy).  The
bolometric luminosity of DOGs with bright 24$\mu$m flux densities is typically
dominated by AGN emission, while the opposite is true for 24$\mu$m faint DOGs
($0.1 \:{\rm mJy} < F_{24\mu {\rm m}} < 0.3 \:$mJy), which are dominated by
star-formation \citep{2008arXiv0808.2816P}.  Additionally, IRAC photometry
shows that the objects in this paper are dominated by a power-law component in
the mid-IR.  The most likely cause of this emission is the presence of warm
dust heated by an AGN \citep{2007ApJ...660..167D}.  

\begin{figure}[!tbp] \epsscale{1.00} \plotone{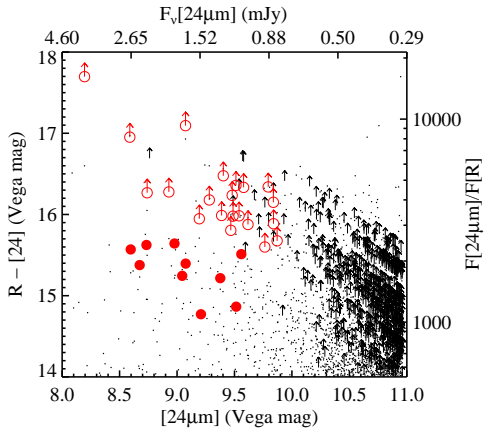}

\caption{$R - [24]$ color vs. 24$\mu$m magnitude distribution for DOGs in
the NDWFS Bo\"{o}tes field.  Bottom and top abscissae show the 24$\mu$m
magnitude and flux density, respectively, and the left and right ordinates show
the color in magnitudes and the $F_{24\mu {\rm m}}/F_R$ flux density ratio,
respectively.  Black dots and upward arrows show the full sample of DOGs, with
and without an $R$-band detection, respectively.  The subsample studied in this
paper is represented by red circles (open symbols show lower limits).
\label{fig:samplecmd}}

\end{figure}

Shallow X-ray coverage of the Bo\"{o}tes field exists and has yielded a full
catalog of X-ray sources
\citep{2005ApJS..161....1M,2005ApJS..161....9K,2006ApJ...641..140B}.  Within a
2$\arcsec$ search radius, two of the DOGs studied in this paper (SST24
J143102.2+325152 and SST24 J143644.2+350627) have a single X-ray counterpart,
and one DOG has two counterparts (SST24 J142644.3+333051).  A full analysis of
the X-ray data is beyond the scope of this paper, but these basic results
suggest that most DOGs are either not strong X-ray emitters or are heavily
obscured.  The latter view is supported both by mid-IR spectral features and
the fact that this subset of 24$\mu$m bright DOGs shows some of the reddest $R
- [24]$ colors of the entire DOG population.  Figure~\ref{fig:samplecmd} shows
the color-magnitude diagram in $R - [24]$ vs.  $[24]$ space for the full DOG
population in Bo\"{o}tes and highlights the subsample of objects studied in
this paper.  

Previous work has shown that objects dominated by a power-law signature in the
mid-IR tend to have AGN indicators in their mid-IR spectra, usually silicate
absorption but no PAH emission
\citep{2006ApJ...653..101W,2008ApJ...675..960P,2008arXiv0801.4578B}.  Indeed,
IRS spectra of these sources have revealed redshifts based on the 9.7$\mu$m
Silicate absorption feature, and all are located at $z \sim 2$.  Of the 31
objects in this sample, 17 have spectra from \citet{2005ApJ...622L.105H}, 2
have spectra from \citet{2006ApJ...651..101W}, and the remaining spectra will
be presented in future work (Higdon et al. in prep.).  Subsequent Keck/NIRSPEC
\citep{2007ApJ...663..204B}, Keck/LRIS, and Keck/DEIMOS spectroscopy has
yielded more precise redshifts for 4 of the DOGs.  The redshift distribution of
the sample studied in this paper compared to the overall distribution of
spectroscopic redshifts for the DOGs from the Bo\"{o}tes field is shown in
Figure~\ref{fig:zdist}.  

\subsection{Observations} \label{sec:obs}

\begin{figure}[!tbp]
\epsscale{1.00}
\plotone{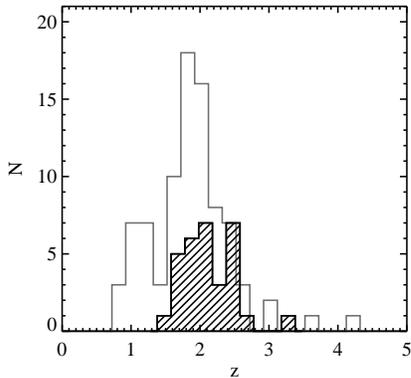}

\caption{Distribution of redshifts for DOGs in the Bo\"{o}tes Field with
spectroscopic redshifts (either from $Spitzer/IRS$ or Keck DEIMOS/LRIS).  The
redshift distribution of the sub-sample of objects studied in this paper is
shown with the hatched histograms and is representative of the full sample of
DOGs.}

\label{fig:zdist}

\end{figure}

The 31 DOGs we study here were observed with {\it HST} from 2006 November to
2008 February.  Nine were imaged in the Wide Field Channel (WFC) mode of ACS
\citep{1998SPIE.3356..234F} before the failure of the instrument.  We have
observed the remaining 22 DOGs with WFPC2 \citep{1994ApJ...435L...3T}.  All 31
DOGs were observed with the NICMOS NIC2 camera.  Table~\ref{tab:observations}
summarizes the details of the observations.  All data were processed using
IRAF\footnote{IRAF is distributed by the National Optical Astronomy
Observatories, which are operated by the Association of Universities for
Research in Astronomy, Inc., under cooperative agreement with the National
Science Foundation.  http://iraf.noao.edu/}.  In the following sections we
provide more details about the processing of the ACS, WFPC2, and NICMOS images
used in this paper.

\begin{deluxetable*}{lllllllll}
\tabletypesize{\tiny} 
\tablecolumns{9}
\tablecaption{Observations}
\tablehead{
\colhead{} & \colhead{} & \colhead{} & \colhead{} &
\multicolumn{2}{c}{Optical Exposures} & 
\multicolumn{2}{c}{Infrared Exposures} \\
\colhead{Source Name} & 
\colhead{RA (J2000)} & 
\colhead{DEC (J2000)} & 
\colhead{$z$} & 
\colhead{Instrument/Filter} & 
\colhead{UT Date} & 
\colhead{Instrument/Filter} & 
\colhead{UT Date}  &
\colhead{ID\tablenotemark{g}}
}
\startdata
SST24 J142538.2+351855  &   14:25:38.155 & +35:18:56.19  &  2.26\tablenotemark{a}    &  WFPC2/F606W & 2007-04-29  & NIC2/F160W    & 2007-06-16  & 19 \\
SST24 J142622.0+345249  &   14:26:22.032 & +34:52:49.69  &  2.00\tablenotemark{c}    &  ACS/F814W   & 2006-11-25  & NIC2/F160W    & 2007-03-13  &  9 \\
SST24 J142626.4+344731  &   14:26:26.538 & +34:47:31.53  &  2.13\tablenotemark{a}    &  WFPC2/F606W & 2007-12-31  & NIC2/F160W    & 2007-01-04  & 16 \\
SST24 J142644.3+333051  &   14:26:44.321 & +33:30:52.20  &  3.312\tablenotemark{d}   &  WFPC2/F606W & 2007-04-10  & NIC2/F160W & 2007-02-25 & 31  \\
SST24 J142645.7+351901  &   14:26:45.701 & +35:19:01.17  &  1.75\tablenotemark{a}    &  WFPC2/F606W & 2007-04-24  & NIC2/F160W    & 2007-05-05  &  2 \\
SST24 J142648.9+332927  &   14:26:48.970 & +33:29:27.56  &  2.00\tablenotemark{c}    &  ACS/F814W   & 2007-01-17  & NIC2/F160W    & 2006-12-19  & 10 \\
SST24 J142653.2+330220  &   14:26:53.285 & +33:02:21.37  &  1.86\tablenotemark{a}    &  ACS/F814W   & 2006-12-29  & NIC2/F160W    & 2007-03-01  &  6 \\
SST24 J142804.1+332135  &   14:28:04.133 & +33:21:34.97  &  2.34\tablenotemark{a}    &  WFPC2/F606W & 2007-04-17  & NIC2/F160W    & 2007-05-01  & 20 \\
SST24 J142924.8+353320  &   14:29:24.811 & +35:33:21.30  &  2.73\tablenotemark{a}    &  WFPC2/F606W & 2007-03-18  & NIC2/F160W    & 2007-02-13  & 30 \\
SST24 J142958.3+322615  &   14:29:58.354 & +32:26:15.17  &  2.64\tablenotemark{a}    &  WFPC2/F606W & 2007-03-14  & NIC2/F160W    & 2007-06-22  & 29 \\
SST24 J143001.9+334538  &   14:30:01.910 & +33:45:38.54  &  2.46\tablenotemark{a}    &  WFPC2/F606W & 2007-04-12  & NIC2/F160W    & 2007-04-28  & 22 \\
SST24 J143025.7+342957  &   14:30:25.764 & +34:29:57.29  &  2.545\tablenotemark{e}   &  WFPC2/F606W & 2007-04-26  & NIC2/F160W & 2007-04-13  & 26 \\
SST24 J143102.2+325152  &   14:31:02.220 & +32:51:52.10  &  2.00\tablenotemark{b}    &  WFPC2/F606W & 2007-04-26  & NIC2/F160W    & 2008-01-07  & 11 \\
SST24 J143109.7+342802  &   14:31:09.823 & +34:28:02.34  &  2.10\tablenotemark{c}    &  WFPC2/F606W & 2007-04-12  & NIC2/F160W    & 2007-01-04  & 13 \\
SST24 J143135.2+325456  &   14:31:35.309 & +32:54:56.84  &  1.48\tablenotemark{c}    &  WFPC2/F606W & 2007-03-21  & NIC2/F160W    & 2007-06-17  &  1 \\
SST24 J143225.3+334716  &   14:32:25.433 & +33:47:16.67  &  2.00\tablenotemark{c}    &  ACS/F814W   & 2006-12-07  & NIC2/F160W    & 2007-06-17  & 12 \\
SST24 J143242.5+342232  &   14:32:42.569 & +34:22:32.23  &  2.16\tablenotemark{a}    &  WFPC2/F606W & 2007-04-19  & NIC2/F160W    & 2007-12-07  & 18 \\
SST24 J143251.8+333536  &   14:32:51.873 & +33:35:35.89  &  1.78\tablenotemark{a}    &  WFPC2/F606W & 2007-03-20  & NIC2/F160W    & 2008-01-14  &  3 \\
SST24 J143312.7+342011  &   14:33:12.734 & +34:20:11.10  &  2.119\tablenotemark{d}   &  WFPC2/F606W & 2007-04-20  & NIC2/F160W & 2007-06-19  & 15 \\
SST24 J143325.8+333736  &   14:33:25.884 & +33:37:36.90  &  1.90\tablenotemark{c}    &  WFPC2/F606W & 2007-05-01  & NIC2/F160W    & 2007-04-20  &  7 \\
SST24 J143358.0+332607  &   14:33:58.077 & +33:26:07.46  &  2.414\tablenotemark{f}   &  ACS/F814W   & 2006-12-10  & NIC2/F160W   & 2008-01-24  & 21 \\
SST24 J143447.7+330230  &   14:34:47.762 & +33:02:30.46  &  1.78\tablenotemark{a}    &  WFPC2/F606W & 2007-03-19  & NIC2/F160W    & 2006-12-23  &  4 \\
SST24 J143504.1+354743  &   14:35:04.166 & +35:47:43.79  &  2.13\tablenotemark{a}    &  WFPC2/F606W & 2007-04-26  & NIC2/F160W    & 2007-01-03  & 17 \\
SST24 J143508.4+334739  &   14:35:08.518 & +33:47:39.44  &  2.10\tablenotemark{c}    &  WFPC2/F606W & 2007-04-12  & NIC2/F160W    & 2007-04-22  & 14 \\
SST24 J143520.7+340418  &   14:35:20.801 & +34:04:18.30  &  1.79\tablenotemark{a}    &  WFPC2/F606W & 2007-03-16  & NIC2/F160W    & 2007-01-03  &  5 \\
SST24 J143523.9+330706  &   14:35:24.005 & +33:07:06.84  &  2.59\tablenotemark{a}    &  ACS/F814W   & 2007-01-01  & NIC2/F160W    & 2007-01-06  & 27 \\
SST24 J143539.3+334159  &   14:35:39.360 & +33:41:59.20  &  2.62\tablenotemark{a}    &  WFPC2/F606W & 2008-05-13  & NIC2/F160W    & 2007-02-15  & 28 \\
SST24 J143545.1+342831  &   14:35:45.137 & +34:28:31.42  &  2.50\tablenotemark{c}    &  ACS/F814W   & 2006-12-06  & NIC2/F160W    & 2007-02-15  & 23 \\
SST24 J143644.2+350627  &   14:36:44.269 & +35:06:27.12  &  1.95\tablenotemark{a}    &  WFPC2/F606W & 2008-01-07  & NIC2/F160W    & 2007-03-14  &   8 \\
SST24 J143725.1+341502  &   14:37:25.186 & +34:15:02.37  &  2.50\tablenotemark{c}    &  ACS/F814W   & 2007-01-07  & NIC2/F160W    & 2007-01-18  & 24 \\
SST24 J143808.3+341016  &   14:38:08.352 & +34:10:15.55  &  2.50\tablenotemark{c}    &  ACS/F814W   & 2006-12-28  & NIC2/F160W    & 2007-02-16  & 25   \\
\tablenotetext{a}{Redshift from {\it Spitzer}/IRS \citep{2005ApJ...622L.105H}}
\tablenotetext{b}{Redshift from {\it Spitzer}/IRS \citep{2006ApJ...653..101W}}
\tablenotetext{c}{Redshift from {\it Spitzer}/IRS (Higdon et al. in prep)}
\tablenotetext{d}{Redshift from Keck NIRSPEC \citep{2007ApJ...663..204B}}
\tablenotetext{e}{Redshift from Keck DEIMOS}
\tablenotetext{f}{Redshift from Keck LRIS}
\tablenotetext{g}{Panel number in Figure~\ref{fig:cutouts1}}
\enddata                                                                          
\label{tab:observations}
\end{deluxetable*}
\subsubsection{ACS} \label{sec:acs} Each DOG was observed over a single orbit
through the F814W filter using a four point dither pattern (ACS-WFC-DITHER-BOX)
with a point spacing of 0.265$\arcsec$, a line spacing of 0.187$\arcsec$ and a
pattern orientation of 20.67$^\circ$.  Total exposure time was
$\approx$2000~sec.  Bias-subtraction and flat-fielding was performed using the
standard ACS pipeline (Pavlovsky et al. 2004).  The MultiDrizzle routine was
used to correct for geometric distortions, perform sky-subtraction, image
registration, cosmic ray rejection and final drizzle combination
\citep{2002hstc.conf..337K}.  We used a square interpolation kernel and set the
output pixel scale at 0.05$\arcsec$~pix$^{-1}$.  

\subsubsection{WFPC2} \label{sec:wfpc2} Following the failure of ACS in the
middle of Cycle~15 observing, the Wide Field Camera CCD 3 of WFPC2 was used to
image the remainder of the DOG population.  For these observations, single-orbit
data through the F606W filter were used to take advantage of WFPC2's superior
sensitivity at this wavelength compared to other WFPC2 filters.  We used a four
point dither pattern (WFPC2-BOX) with a point and line spacing of 0.559$\arcsec$
and a pattern orientation of 26.6$^\circ$.  Total exposure time was
$\approx$1600~sec.  The standard WFPC2 pipeline system was used to
bias-subtract, dark-subtract, and flat-field the images (Mobasher et al., 2002).
MultiDrizzle was then used to correct for geometric distortions, perform
sky-subtraction, image registration, cosmic ray rejection and final drizzle
combination \citep{2002hstc.conf..337K}.  We used a square interpolation kernel
and output pixel scale of 0.045$\arcsec$~pix$^{-1}$, leading to a per-pixel
exposure time of $\approx$340~sec.  Due to the irregular performance of WF4 and
the PC CCDs, we have restricted our analysis to the WF2 and WF3 CCDs.

\subsubsection{NICMOS} \label{sec:nicmos} Single orbit data of the DOGs were
acquired with NIC2 and the F160W filter.  We used a two-point dither pattern
(NIC-SPIRAL-DITH) with a point spacing of 0.637$\arcsec$.  Total exposure time
was $\approx$2600~s.  To reduce the data, we followed the standard reduction
process outlined in the NICMOS data handbook (McLaughlin \& Wikland 2007).  We
used the IRAF routine {\tt nicpipe} to pre-process the data, followed by the
{\tt biaseq} task to correct for non-linear bias drifts and spatial bias jumps.
We then used {\tt nicpipe} a second time to do flat-fielding and initial
cosmic-ray removal.  The IRAF task {\tt pedsky} was used to fit for the sky
level and the quadrant-dependent residual bias.  Significant residual
background variation remained after this standard reduction process.  To
minimize these residuals, we constructed a normalized, object-masked median sky
image based on all of our NIC2 science frames.  This sky image was then scaled
by a constant factor and subtracted from each science image.  The scaling
factor was computed by minimizing the residual of the difference between the
masked science image and the scaled sky image.  Mosaicing of the dithered
exposures was performed using {\tt calnicb} in IRAF, resulting in a pixel scale
of 0.075$\arcsec$~pix$^{-1}$.

\subsection{Astrometry}\label{sec:astro}

Each ACS/WFPC2 and NICMOS image is aligned to the reference frame of the NDWFS,
which itself is tied to the USNO A-2 catalog.  We first run Source Extractor
\citep[SExtractor,][]{1996A&AS..117..393B} on a cutout of the $I$-band NDWFS
corresponding to the appropriate ACS/WFPC2 Field Of View (FOV) to generate a
list of comparison objects.  The IRAF task {\tt wcsctran} is used to convert
this list into pixel coordinates on the ACS/WFPC2 image.  Another IRAF task,
{\tt imcentroid}, is used to improve the accuracy of the pixel coordinates.
Finally, the IRAF task {\tt ccmap} applies a first order fit to correct the
zero point of the astrometry and update the appropriate WCS information in the
header of the ACS/WFPC2 image.  This aligned ACS/WFPC2 image is then used as
the reference frame for correcting the astrometry of the NICMOS image and the
IRAC images \citep[since the IRAC images of the Bo\"{o}tes Field are not tied
to the USNO A-2 catalog, but instead to the 2$\mu$m All-Sky Survey frames,
see][]{2004ApJS..154...48E}.  Using the properly aligned, multi-wavelength
dataset, identifying the proper counterpart to the MIPS source is relatively
straightforward, since inspection of the four IRAC channels reveals a single
source associated with the 24$\mu$m emission for all but one source (this
source is undetected in all four IRAC channels).  The absolute uncertainty in
the centroid of the IRAC 3.6$\mu$m emission ranges from 0$\farcs$3-0$\farcs$5.

\subsection{Photometry}\label{sec:photo}

We perform 2$\arcsec$ diameter aperture photometry on each DOG in both the
rest-optical and rest-UV, choosing the center of the aperture to be located at
the peak flux pixel in the NICMOS images.  We remove foreground and background
objects using SExtractor (see Section~\ref{sec:nonpar}) and calculate the sky
level using an annulus with an inner diameter of 2$\arcsec$ and a width of
1$\arcsec$.  We found that in some cases (particularly those NICMOS images where
significant residual non-linearities remained), the flux density radial profile
did not flatten at large radii.  When this occurred, we determined the
appropriate sky value by trial-and-error.   We computed the background level
and photometric uncertainty by measuring the sigma-clipped mean and RMS of
fluxes measured in $N$ $2\arcsec$ diameter apertures, where $N \approx 10$ and
$N \approx 50$ for the NICMOS and ACS/WFPC2 images, respectively.

We compute 4$\arcsec$ diameter aperture photometry in the NDWFS $B_W$, $R$, and
$I$ images centered on the IRAC 3.6 $\mu$m centroid of emission.  Sky
background levels were computed in a 3$\arcsec$ wide annulus with an inner
diameter of 4$\arcsec$.  Limiting magnitudes were determined by measuring the
flux within a 4$\arcsec$ aperture at several sourceless locations near the DOG
and computing the rms variation of the flux values.

We verified the accuracy of our ACS and WFPC2 photometric zeropoints by
comparing well-detected sources common to both our {\it HST} and NDWFS imaging.
For our ACS/F814W observations, we compared to the NDWFS $I$-band imaging and
found negligible offsets (-0.03 $\pm$ 0.10 magnitudes).  For our WFPC2/F606W
observations, we compared to the NDWFS $R$-band (after correcting for color
terms due to the dissimilarity of the $R$ and F606W filter bandpasses) and
again found negligible offsets (0.05 $\pm$ 0.15 magnitudes).

\subsection{Images of DOGs}\label{sec:galim}

Figure~\ref{fig:cutouts1} shows $2\arcsec \times 2\arcsec$ cutout images of the
DOGs in order of increasing redshift.  Each cutout is centered roughly on the
centroid of emission as seen in the NICMOS image.  A red plus sign shows the
centroid of IRAC 3.6$\mu$m emission and is sized to represent the 1-$\sigma$
uncertainty in the position, which includes independent contributions from the
centroiding error on the 3.6$\mu$m emission ($\approx$0$\farcs$2-0$\farcs$4,
depending on S/N), the relative astrometric calibration uncertainty within the
3.6$\mu$m map ($\approx$0$\farcs$2), and the uncertainty in tying the 3.6$\mu$m
map to the {\it HST} images ($\approx$0$\farcs$1).  The 1$\sigma$ rms offset
between IRAC and NICMOS centroids of the sample is 0$\farcs$2.  In most cases,
the offset in centroids is negligible, but those cases where it is not are
associated with faint 3.6$\mu$m emission (when the absolute astrometric
uncertainty may be as large as 0\farcs5).  This suggests there is no
significant offset between the near-IR and mid-IR centroids, although we note
that we cannot rule out offsets at the $< 1$~kpc scale.

\begin{figure*}[!tbp]
\epsscale{1.00}
\plotone{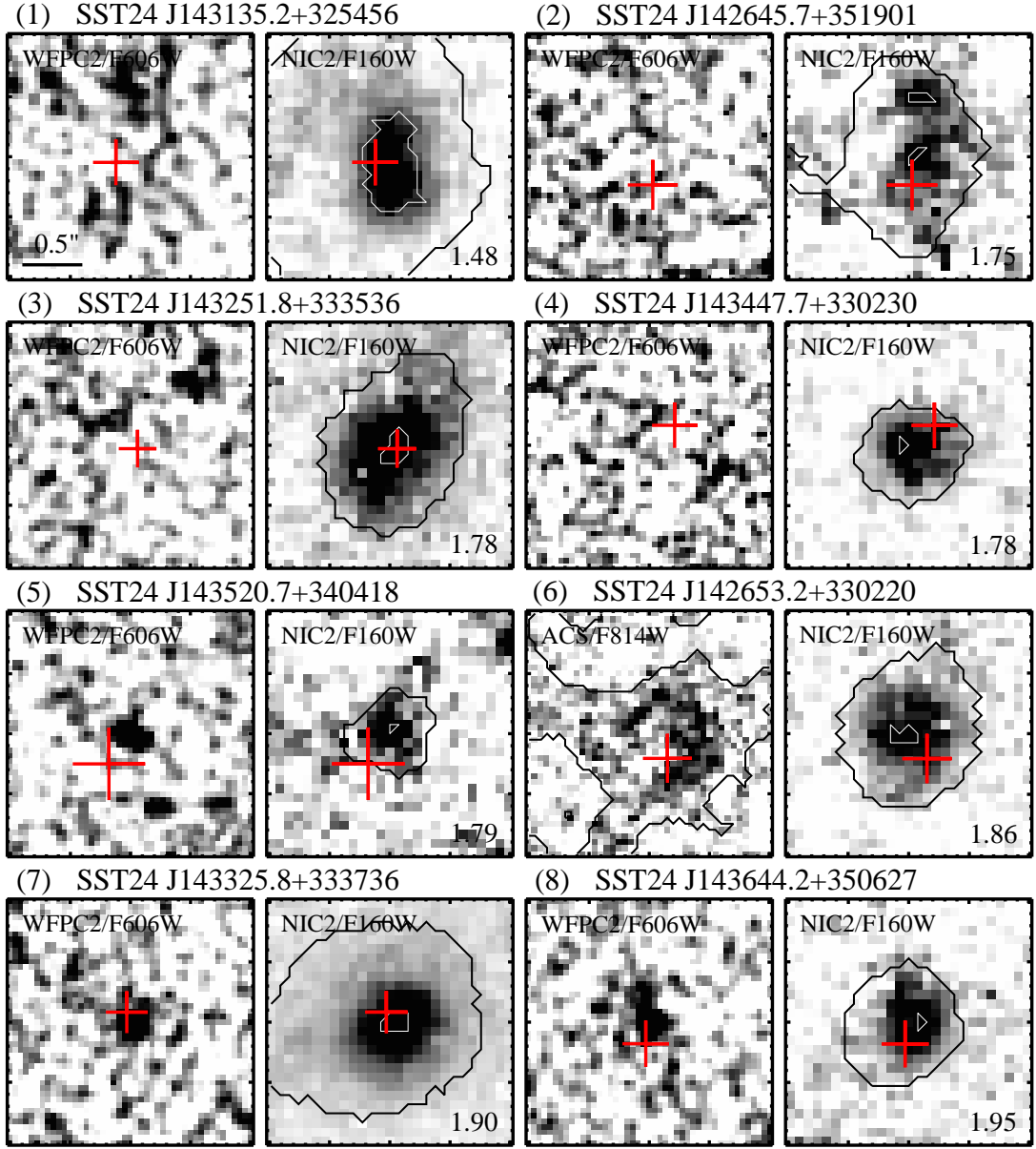}

\caption{Cutouts of the 31 DOGs observed by {\it HST}, shown with a linear
stretch.  Columns 1 and 3 are the rest-UV images from either ACS F814W or WFPC2
F606W.  Columns 2 and 4 are the rest-optical images from NIC2 F160W.  Each
cutout is 2$\arcsec$ on a side and is oriented north up and east left.  The
objects are arranged in order of increasing redshift, and the redshift is
printed in the lower right corner of each NICMOS image.  A red cross denotes
the position and 1-$\sigma$ uncertainty in the centroid of the IRAC 3.6$\mu$m
emission.  In NICMOS images where the S/N per pixel is greater than 2, white
contours outline the brightest 20\% pixels, and black contours show the outline
of the segmentation map used in measuring the non-parametric morphologies.}

\label{fig:cutouts1}
\addtocounter{figure}{-1}
\end{figure*}

\begin{figure*}[!tbp]
\epsscale{1.00}
\plotone{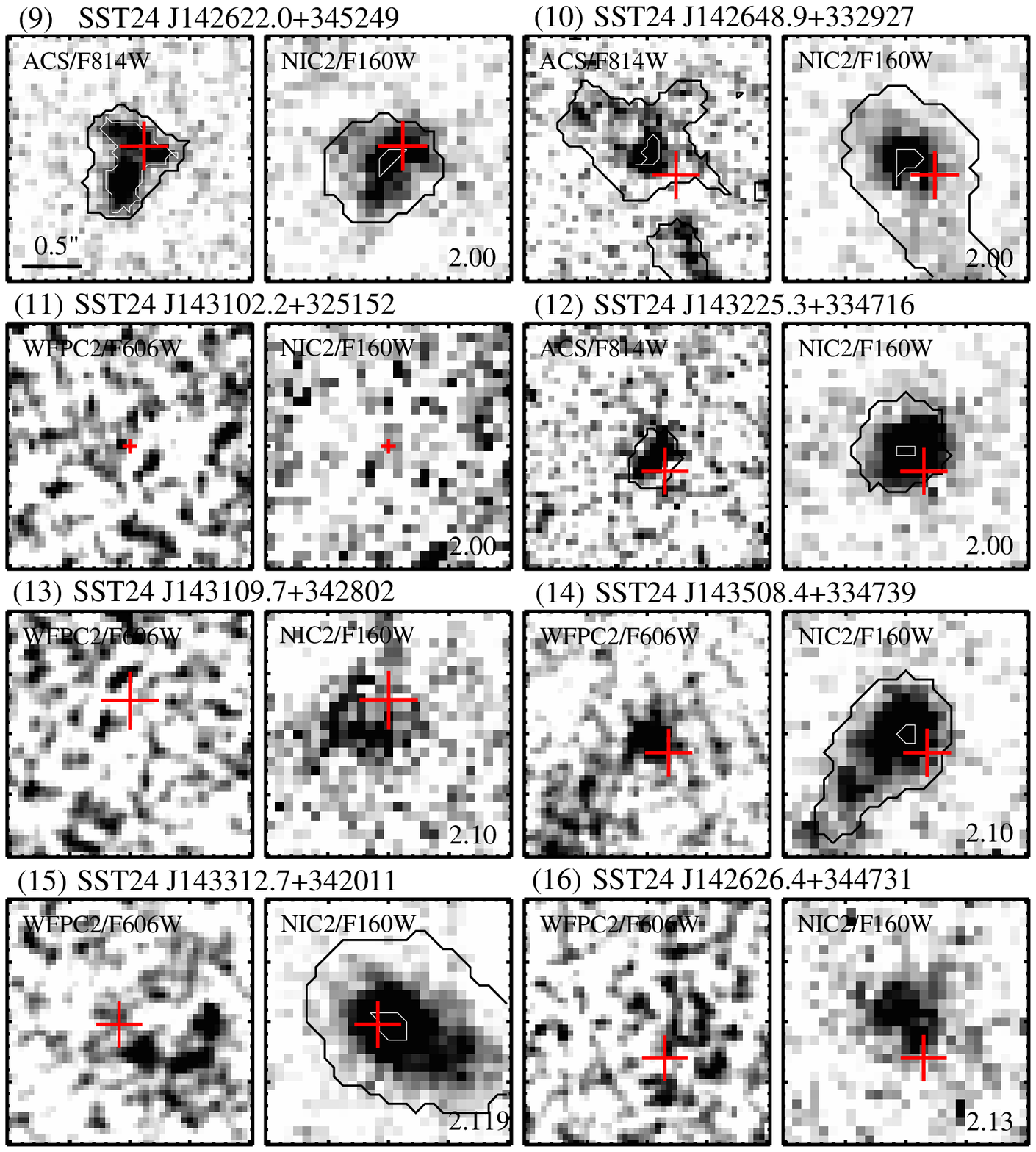}

\caption{Continued.}

\label{fig:cutouts2}
\addtocounter{figure}{-1}
\end{figure*}

\begin{figure*}[!tbp]
\epsscale{1.00}
\plotone{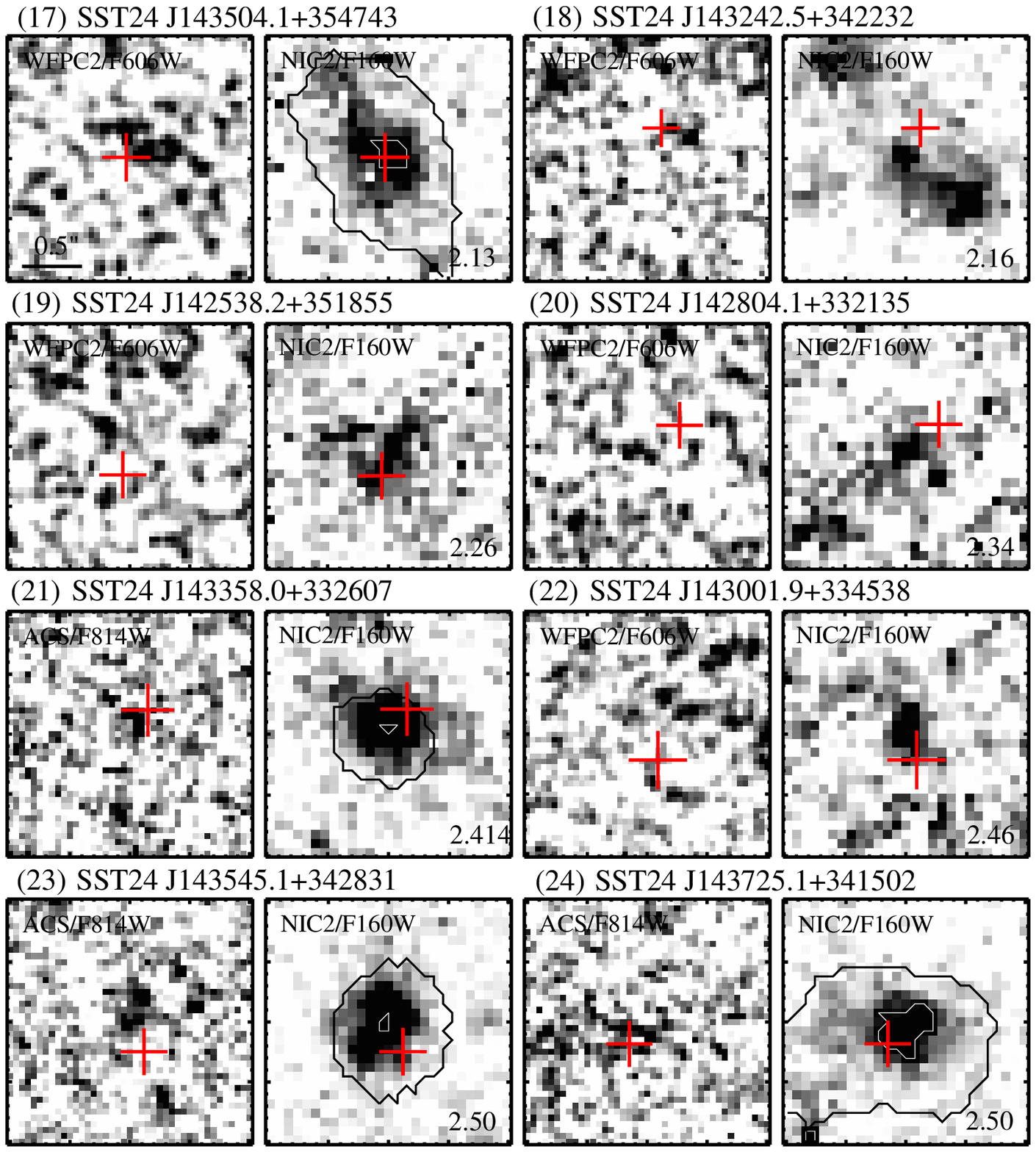}

\caption{Continued.}

\label{fig:cutouts3}
\addtocounter{figure}{-1}
\end{figure*}

\begin{figure*}[!tbp]
\epsscale{1.00}
\plotone{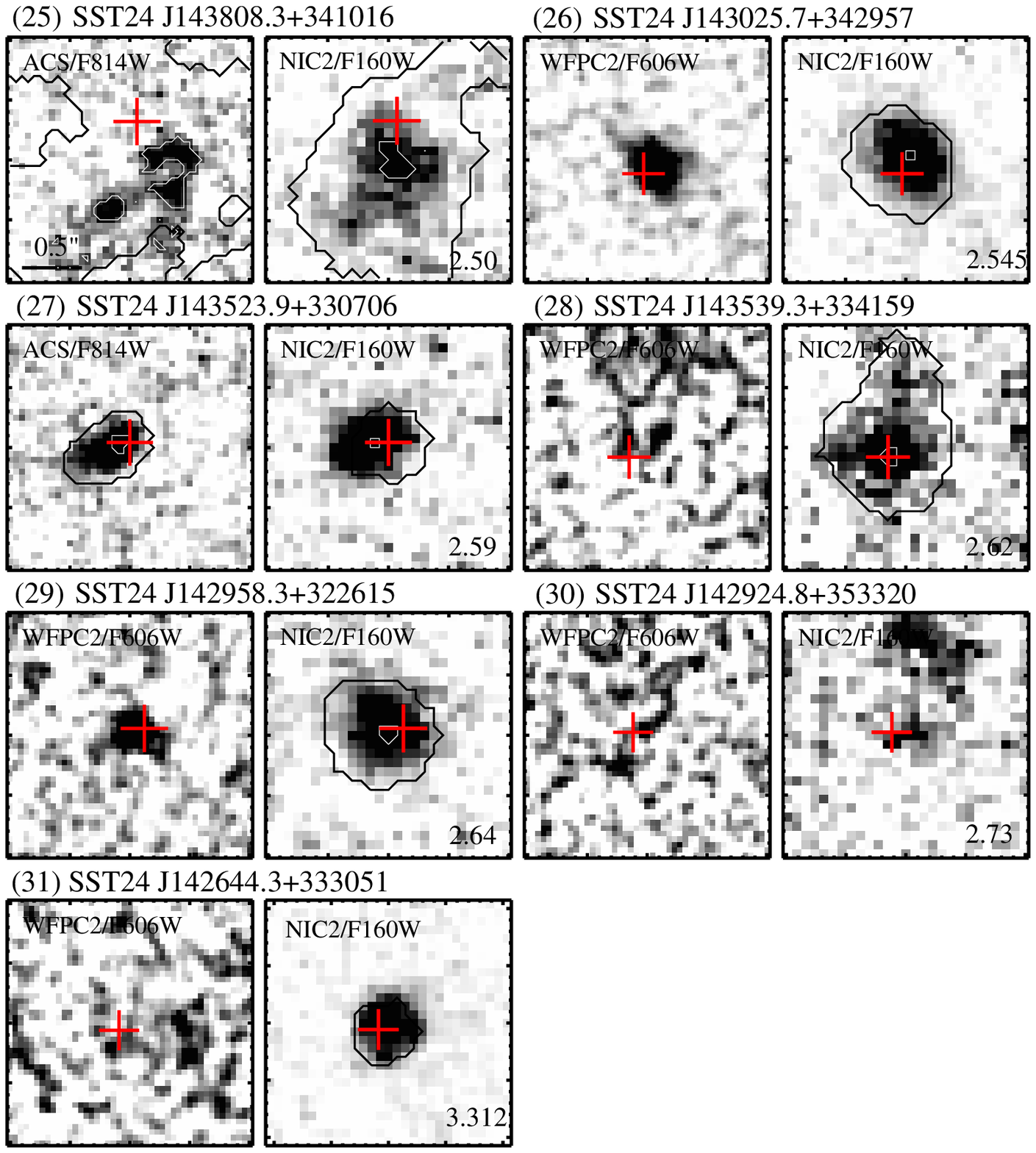}

\caption{Continued.}

\label{fig:cutouts4}

\end{figure*}

The DOGs exhibit a wide range of morphologies, with most being well-resolved.
Only one object (SST24 J142644.3+333051) shows strong Airy rings and is clearly an
unresolved point source.  Here we give a brief qualitative desription of the
morphology of each object.

{\bf (1) SST24 J143135.2+325456:}  F606W: Weak detection.  F160W:
Large-scale emission with a faint tail extending northeast.

{\bf (2) SST24 J142645.7+351901:}  F606W: No significant detection.  F160W: Two
compact, resolved components separated by $\approx$0\farcs5.  

{\bf (3) SST24 J143251.8+333536:}  F606W: Faint emission slightly NE of NICMOS
centroid; possible second source $\sim$0$\farcs$5 northwest (NW) of NICMOS
centroid.  F160W: Extended object; possible point source contamination.  

{\bf (4) SST24 J143447.7+330230:}  F606W: No significant detection.  F160W: Irregular,
diffuse object.

{\bf (5) SST24 J143520.7+340418:}  F606W: Compact, resolved object.  F160W: Very
compact object, but no evidence for Airy rings.  

{\bf (6) SST24 J142653.2+330220:}  F814W: Large scale, irregular, and diffuse
emission. F160W: Large-scale, irregular, and diffuse, but with bright compact
nuclear component.  

{\bf (7) SST24 J143325.8+333736:}  F606W: Compact, resolved object. F160W: Bright
compact and extended components.  

{\bf (8) SST24 J143644.2+350627:}  F606W: Compact, resolved object. F160W:
Extended object; possible point source contamination.  This object has a
counterpart in the XBo\"otes catalog \citep{2006ApJ...641..140B}.

{\bf (9) SST24 J142622.0+345249:}  F814W: Four compact, resolved clumps spread in a
`T' shape with no visible central component. F160W: Similar irregular `T'
shape, but components are not as distinct.  NE component is bluer than other
components.

{\bf (10) SST24 J142648.9+332927:}  F814W: Compact, resolved object.  Chain of sources
extends towards southwest.  One of these sources is within 1$\arcsec$ of the
DOG and is included in the photometric and morphological measurements, since
there is no clear evidence to suggest it is not associated with the system.
F160W: Similar compact, resolved object; possible point source contamination.

{\bf (11) SST24 J143102.2+325152:}  F606W: No significant detection. F160W: No
significant detection.  This object is also not detected in any of the IRAC
images, but does have a counterpart in the XBo\"otes catalog
\citep{2006ApJ...641..140B}.

{\bf (12) SST24 J143225.3+334716:}  F814W: Compact, irregular object. F160W: Compact,
resolved object; possible point source contamination.  

{\bf (13) SST24 J143109.7+342802:}  F606W: No significant detection. F160W: Irregular,
diffuse object.

{\bf (14) SST24 J143508.4+334739:}  F606W: Compact, resolved, and irregular central
component with tail of emission to southeast (SE). F160W: Very similar, but
central component is stronger relative to tail.

{\bf (15) SST24 J143312.7+342011:}  F606W: Four compact components in a semi-circle
offset from the centroid of NICMOS emission. F160W: Extended object; possible
point source contamination.

{\bf (16) SST24 J142626.4+344731:}  F606W: No significant detection. F160W: Irregular,
diffuse object.

{\bf (17) SST24 J143504.1+354743:}  F606W: Faint source barely detected. F160W:
Irregular, extended object; possible point source contamination.

{\bf (18) SST24 J143242.5+342232:}  F606W: No significant detection. F160W: Faint,
compact component near 3.6$\mu$m centroid with emission leading to second,
brighter peak $\sim$0$\farcs$5 to southwest (SW).  Possible multiple-component
system.

{\bf (19) SST24 J142538.0+332607:}  F606W: No significant detection. F160W: Irregular
objected elongated in NW-SE direction.

{\bf (20) SST24 J142804.1+332135:}  F606W: No significant detection. F160W: Faint,
irregular object.

{\bf (21) SST24 J143358.0+332607:}  F606W: Faint, irregular object. F160W: Extended
object with possible point source contamination.

{\bf (22) SST24 J143001.9+334538:}  F606W: No significant detection. F160W: Faint,
irregular object.

{\bf (23) SST24 J143545.1+342831:}  F814W: Faint, compact irregular objects; possibly
two component system. F160W: Extended emission; possible point source
contamination.

{\bf (24) SST24 J143725.1+341502:}  F814W: Very faint, low surface brightness feature
extending to east. F160W: Diffuse emission with compact object; faint Airy
ring present.

{\bf (25) SST24 J143808.3+341016:}  F814W: Faint, compact, resolved components offset
from centroid of NIC2 emission. F160W: Compact, central component with
extension to SE overlapping ACS emission centroid.

{\bf (26) SST24 J143025.7+342957:}  F606W: Compact, resolved object. F160W: Compact,
resolved object; possible point source contamination.

{\bf (27) SST24 J143523.9+330706:}  F814W: Compact, resolved object with tail of
emission to SE; possible point source contamination. F160W: Compact,
resolved object; possible point source contamination.

{\bf (28) SST24 J143539.3+334159:}  F606W: Possible faint diffuse emission N
of NIC2 centroid. F160W: Compact, resolved object; possible point source
contamination; possible tail of emission towards N.

{\bf (29) SST24 J142958.3+322615:}  F606W: Compact resolved object; possible point
source contamination. F160W: Extended object with bright nuclear source;
possible point source contamination.

{\bf (30) SST24 J142924.8+353320:}  F606W: No significant detection. F160W: Very
compact, irregular, faint object near IRAC 3.6$\mu$m centroid.  Larger,
brighter object $\sim$0\farcs8 to north.

{\bf (31) SST24 J142644.3+333051:}  F606W: Weak detection. F160W: Dominated by
point source emission; clear Airy ring.  This source has two X-ray counterparts
in the XBo\"otes catalog \citep{2006ApJ...641..140B}.

\section{Methodology: Morphological Analysis}\label{sec:morphmeth}

We undertook three different, complementary approaches to analyzing the
morphology of the DOGs in our sample: a visual classification experiment,
multi-component GALFIT modeling, and non-parametric quantification.  In this
section, we describe the details of our methodology.  The results are described
in section~\ref{sec:results}.

\subsection{Visual Classification} \label{sec:visualmeth}

We first undertook a visual classification of the DOGs by conducting the
following experiment:  for each ACS/WFPC2 image, we generate a
5$\arcsec$x5$\arcsec$ cutout image of both the DOG and 14 other randomly
selected galaxies in the same FOV with the same magnitude range as our DOG
sample.  Eight of the coauthors then classified all 15 galaxies in each FOV
(the DOG was not identified), placing them into one of the following bins:
Elliptical/Compact (E/C), Disk, Irregular/Multi-component, Irregular/Diffuse,
or Too Faint to Tell.  In practice, since these galaxies are selected to be
faint in the optical and generally have low S/N, we group together the E/C and
Disk categories into a ``Regular'' bin and the two irregular categories into an
``Irregular'' bin.  This results in a total of 3600 independent
classifications, of which 240 pertain to the DOGs.  In an effort to explore the
robustness of our results, we flag and remove from the sample all objects where
fewer than six classifers were in agreement.  This has the effect of reducing
the fraction of ``Too Faint To Tell'' responses, but the ratio of the Regular
to Irregular classifications changes by less than 15\%.  A similar experiment
was done on the NICMOS images, without the control sample, since the NIC2 FOV
is so small.

Interpretation of the results of our visual classification analysis is hampered
by low S/N (in the case of the ACS/WFPC2 images) or the lack of a control
sample (in the case of the NICMOS images), so we forego a detailed analysis and
instead present the mode of the classifications for each DOG along with an
indication of whether the eight coauthors were in general agreement in our
morpholical results table in section~\ref{sec:nonpar}.  This is useful as a
qualititative assessment of the morphology for comparison with the more
quantitative methods discussed below.

\subsection{GALFIT Modeling}\label{sec:parmeth}

In many of the NICMOS images, there is a compact component that is not seen in
the corresponding ACS/WFPC2 image, implying there is significant obscuration in
the central region of many of the DOGs.  In this section, we describe the
method we use to explore the degree to which each DOG is dominated by a
central, unresolved component.  Our tool in this effort is GALFIT
\citep{2002AJ....124..266P}, which uses a 2-dimensional $\chi^2$ minimization
to search the parameter space of a set of predefined functions and identify the
parameters that best describe the observed 2-D profile.  

Because the DOGs are small and have low S/N compared to more typical
applications of GALFIT, we restrict the size of the fitting region to be
41$\times$41 pixels (corresponding to an angular and physical size of
3$\arcsec$ and $\approx$24~kpc, respectively) and include the minimum necessary
components in our model.  For a variety of reasons, we expect AGN to be
important contributors to the emitted radiation from these sources.  Therefore,
we model the observed emission with three components which are described by a
total of 10 free parameters.  The number of degrees of freedom, $N_{\rm DOF}$,
is calculated as the number of pixels in the image being modeled minus the
number of free parameters.  This implies that the maximum $N_{\rm DOF}$ is
1671.  Those cases where $N_{\rm DOF} < 1671$ are associated with images where
pixels were masked out because they were associated with residual instrumental
noise and prevented convergence with GALFIT.  We note that because NIC2 is a
Nyquist-sampled array (0.075$\arcsec$~pix$^{-1}$ compared to 0.16$\arcsec$ FWHM
beam), the pixels in our image may not be completely independent.  As a result,
the $\chi^2_\nu$ values should be interpreted in a relative sense rather than
an absolute one.

The first element in our GALFIT model is a sky component whose amplitude is
chosen to obtain flat radial profiles at large radii and is not allowed to
vary.  The second is an instrumental PSF generated from the TinyTim software
(Krist and Hook 2004), which can simulate a PSF for NICMOS, WFPC2, and ACS.
For the NICMOS and WFPC2 images, the DOG is positioned in nearly the same spot
on the camera.  In the case of WFPC2 this is pixel (400,400) of chip 3 and
pixel (155, 164) for NICMOS.  Meanwhile, a different region of the ACS camera
is used for each DOG.  Therefore, we generate a unique PSF at each position on
the appropriate chip in which a DOG is observed.  We use a red power-law
spectrum ($F_\nu \propto \nu^{-2}$) as the object spectrum.  The PSF is
computed out to a size of 3.0$\arcsec$, and for the WFPC2 PSF we oversample by
a factor of 2 to match the pixel scale of the drizzled WFPC2 images.

The final component is a S\'{e}rsic profile \citep{1968adga.book.....S} where
the surface brightness scales with radius as exp[$-\kappa ( (r/R_{\rm
eff})^{1/n}-1)$], where $\kappa$ is chosen such that half of the flux falls
within $R_{\rm eff}$.  We attempted to place as few constraints as possible so
as to optimize the measurement of the extended flux (i.e., non-point source
component).  However, in certain cases, the S\'{e}rsic index had to be
constrained to be positive to ensure convergence on a realistic solution.  For
the NICMOS images, we used the uncertainty image output by {\tt calnicb} as the
error image required by GALFIT to perform a proper $\chi^2$ minimization.  The
TinyTim NIC2 PSF is convolved with the S\'ersic profile prior to performing the
$\chi^2$ minimization.  The initial guesses of the magnitude, half-light
radius, position angle, and ellipticity were determined from the output values
from SExtractor.  Varying the initial guesses within reasonable values (e.g.,
plus or minus two pixels for the half-light radius) yielded no significant
change in the best-fit model parameters.  We used the NICMOS centroid as the
initial guess for the (x,y) position of both the PSF and extended components,
but in a few cases these guesses had to be modified by 1-2 pixels in order to
result in convergence.  

We note that we tested two-component models as well (single component S\'ersic
profile plus sky background) and found larger reduced $\chi^2$ values,
especially when the point source fraction in our three-component model was
large (see further discussion in section~\ref{sec:galfit}).  In cases where the
point source fraction was small, the two-component model had similar parameter
values as the three-component model, as we would expect.

It is important to note here that NIC2 cannot spatially resolve objects smaller
than 1.3~kpc at $z \approx 2$.  This limit is large enough to encompass a
compact stellar bulge as well as an active galactic nucleus, implying that we
cannot, from these data alone, distinguish between these two possibilities as
to the nature of the central, unresolved component.

After the best-fit parameters are found in the NICMOS image, we run GALFIT with
a simplified model on the DOGs in the ACS/WFPC2 images.  The primary
simplification is to fix the position of the PSF based on the best-fit NICMOS
PSF location (allowing up to 2 pixel wiggle room to account for astrometric
uncertainties, which can be as large as 0\farcs1).  In many cases, GALFIT
required an upper limit to be placed on the magnitude of the PSF component in
order to reach convergence.  We choose to use the magnitude of a point source
detected at the 2-$\sigma$ level for this upper limit.  We note that our
S\'{e}rsic profile model for the extended DOG flux is not representative of the
rest-UV morphology of many of the DOGs (i.e., the reduced $\chi^2$ values are
large), but it does adequately recover their total flux.

\subsection{Non-parametric Classification} \label{sec:nonparmeth}

The Gini coefficient ($G$) and $M_{20}$ parameter are known to be reliable
tools for the characterization of faint-object morphologies
\citep{2004AJ....128..163L}.  $G$ was originally created to measure how evenly
the wealth in a society is distributed \citep{glasser1962}.  Recently,
\citet{2003ApJ...588..218A} and \citet{2004AJ....128..163L} applied this method
to aid in the classification of galaxies, with $G$ defined such that low (high)
values imply an equal (un-equal) distribution of flux.  $M_{20}$ is the
logarithm of the second-order moment of the brightest 20\% of the galaxy's
flux, normalized by the total second-order moment \citep{2004AJ....128..163L}.
This means that higher values of $M_{20}$ imply multiple bright clumps offset
from the second-order moment center.  Lower values, on the other hand, suggest
a system dominated by a central component.

Prior to computing $G$ or $M_{20}$, we first generate a catalog of objects
using SExtractor.  We use a detection threshold of 1.5$\sigma$ (corresponding
to 24.5~mag~arcsec$^{-2}$) and a minimum detection area of 15~pixels.  The
center of the image as well as the ellipticity and position angle computed by
SExtractor are used as inputs for computing morphological measures.  In
addition, we use catalog sources selected to have magnitudes within the range
of all 24 DOGs analyzed in this paper as a ``field'' galaxy sample  for
comparison to DOGs.  

Much of the methodology in this section relies on morphology code written by J.
Lotz and described in detail in \citet{2004AJ....128..163L}.  Here, we
summarize the relevant information.  Postage stamps of each object in the
SExtractor catalog (and the associated segmentation map) are created with
foreground/background objects masked out.  Using a small region of the cutout
devoid of sources, a sky value is computed and subtracted from the postage
stamp.  Next, we determine which pixels in each postage stamp belong to the
galaxy and which do not.  Since the isophotal-based segmentation map produced
by SExtractor is subject to the effects of surface brightness dimming at high
redshift, we use a segmentation map based on the mean surface brightness at the
Petrosian radius $\mu(R_p)$.  Pixels with surface brightness above $\mu(R_p)$
are assigned to the galaxy while those below it are not.  We define $R_p$ as
the radius at which the ratio of the surface brightness at $R_p$ to the mean
surface brightness within $R_p$ is equal to 0.2.  

Using the new segmentation map, we recompute the galaxy's center by minimizing
the total second-order moment of the flux.  A new value of $R_p$ is then
computed and a revised segmentation map is used to calculate $G$ and $M_{20}$.
Finally the morphology code produces an average S/N per pixel value using the
pixels in the revised segmentation map \citep[Eqs.~1 through 5
in][]{2004AJ....128..163L}.

One of the most common methods of characterizing galaxy morphologies in the
literature is to measure the concentration index $C$
\citep{1994ApJ...432...75A}, the rotational asymmetry $A$
\citep{1995ApJ...451L...1S}, and the residual clumpiness, $S$
\citep{2003ApJS..147....1C}.  Given sufficiently high S/N and spatial
resolution, the $CAS$ system has had demonstrated success in measuring
morphological parameters and identifying mergers at low
\citep{2003ApJS..147....1C} and high redshift \citep{2008MNRAS.386..909C}.
Unfortunately, the objects in our sample do not meet simultaneously the S/N and
spatial resolution requirements to be reliably placed in $CAS$ space.  Because
computation of $A$ and $S$ involves differencing two images, the necessary
per-pixel S/N to measure these parameters reliably is twice as high as those
that do not involve subtracting images.  We find per-pixel S/N values ranging
from $\sim$2-5 for the DOGs, whereas reliable measurements of $A$ and $S$
require S/N$\geq$5 \citep{2004AJ....128..163L}.  In principle, the data are of
sufficient quality to measure $C$ (see Tab.~\ref{tab:morph}), but in practice
we find that the inherent assumption of circular symmetry does not apply well
to the DOGs, making the interpretation of $C$ values difficult.  

\section{Results}\label{sec:results}

\subsection{Photometry}\label{sec:photres}

\begin{figure*}[!tbp]
\epsscale{1.2}
\plotone{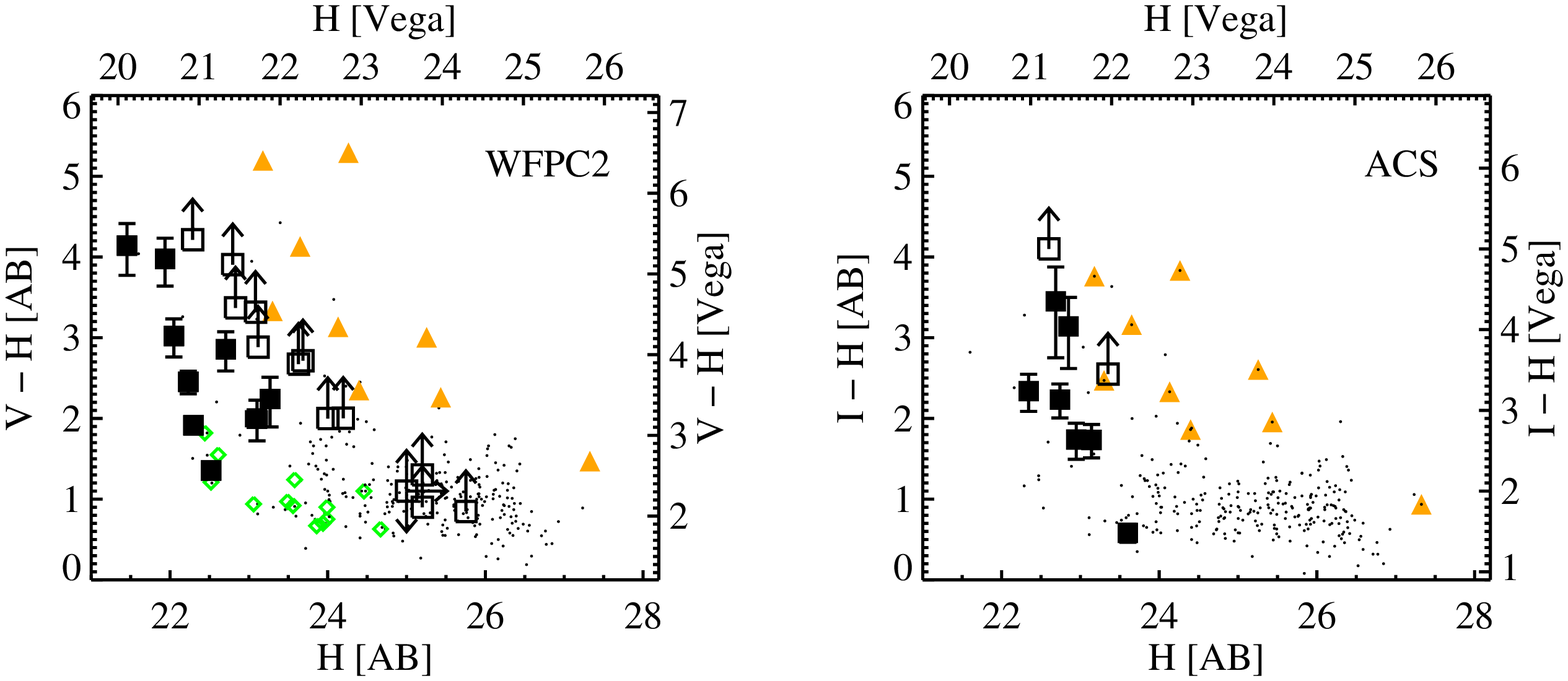}

\caption{ Color-magnitude diagram for DOGs.  {\it Left:} $V - H$ vs. $H$ for
DOGs observed by WFPC2 (filled black squares show detections, open black
squares show lower limits).  Galaxies spanning the redshift range $1.5 < z <
2.5$ in the HDF-N (Papovich, personal communication) and HDF-S
\citep{2003AJ....125.1107L} are shown with black dots.  Bright LBGs from the
HDF-N are shown with green diamonds \citep{2001ApJ...559..620P}.  DRGs in the
HDF-S are represented by filled orange triangles.  {\it Right:} $I - H$ vs. $H$
for DOGs observed by ACS.  Symbols are the same as in left panel.  }
\label{fig:imh}

\end{figure*}

In Figure~\ref{fig:imh}, we show the color-magnitude diagram for DOGs and a
sample of galaxies in the HDF whose photometric redshifts are comparable to
DOGs ($1.5 < z_{\rm phot} < 2.5$).  For DOGs where the measured flux is below
the 2$\sigma$ detection limit, we use an open plotting symbol and an upward
pointing arrow.  DOGs range in $H$-band magnitude from 21.93 to 25.1 AB mags.
In both $V - H$ and $I - H$, DOGs are redder than a typical high-$z$ galaxy by
0.2-3 AB mags.  In particular, the LBGs from the HDF-N
\citep{2001ApJ...559..620P} are comparably bright in $H$, but fainter in $V$ by
$\approx$2 AB mags than DOGs.  There is substantial overlap between the colors
of DOGs and DRGs, suggesting that we might expect to see similarities in the
morphologies between these two populations.  Table~\ref{tab:sources} summarizes
the photometric information derived from the NDWFS and the {\it HST} imaging.

\begin{deluxetable*}{lccccccccc}
\tabletypesize{\tiny} 
\tablecolumns{10}
\tablewidth{450pt}
\tablecaption{Photometric Properties\tablenotemark{a}\tablenotemark{b}}
\tablehead{
\colhead{Source Name} & 
\colhead{$B_W$} & 
\colhead{$R$} & 
\colhead{$I$} & 
\colhead{$V$ (F606W)} & 
\colhead{$I$ (F814W)} & 
\colhead{$H$ (F160W)} & 
\colhead{$F_{24}$~(mJy)} &
\colhead{$R - [24]$\tablenotemark{c}}
}
\startdata
SST24 J142538.2+351855 & $>$ 26.6 & $>$ 25.9 & $>$ 25.5 &             $>$ 26.0 & --- & 24.0$\pm$0.1         & 0.85$\pm$0.05 & $>$ 16.1 \\
SST24 J142622.0+345249 & 24.5$\pm$0.1 & 24.5$\pm$0.3 & 24.0$\pm$0.3 & --- & 24.18$\pm$0.06 & 23.6$\pm$0.1   & 1.29$\pm$0.05 &     15.2 \\
SST24 J142626.4+344731 & $>$ 26.6 & $>$ 25.4 & $>$ 25.2 &             $>$ 26.4 & --- & 23.7$\pm$0.1         & 1.17$\pm$0.04 & $>$ 16.0 \\
SST24 J142644.3+333051 & $>$ 26.5 & 24.3$\pm$0.2 & 24.3$\pm$0.2 &     25.9$\pm$0.3 & --- & 21.93$\pm$0.02   & 1.14$\pm$0.04 &     14.9 \\
SST24 J142645.7+351901 & $>$ 26.6 & $>$ 25.8 & 24.5$\pm$0.3 &         $>$ 26.2 & --- & 23.31$\pm$0.09       & 1.14$\pm$0.05 & $>$ 16.3 \\
SST24 J142648.9+332927 & 25.1$\pm$0.2 & $>$ 25.0 & 24.1$\pm$0.1 &     --- & 24.9$\pm$0.2 & 23.3$\pm$0.1     & 2.33$\pm$0.07 & $>$ 16.3 \\
SST24 J142653.2+330220 & $>$ 26.6 & $>$ 26.1 & 24.7$\pm$0.3 &         --- & 25.0$\pm$0.2 & 22.7$\pm$0.1     & 0.88$\pm$0.05 & $>$ 16.3 \\
SST24 J142804.1+332135 & $>$ 26.4 & $>$ 25.7 & $>$ 25.3 &             $>$ 26.6 & --- & 25.1$\pm$0.5         & 0.84$\pm$0.03 & $>$ 15.9 \\
SST24 J142924.8+353320 & $>$ 26.6 & $>$ 25.4 & $>$ 24.9 &             $>$ 26.1 & --- & 24.7$\pm$0.3         & 1.04$\pm$0.05 & $>$ 15.9 \\
SST24 J142958.3+322615 & 25.6$\pm$0.1 & $>$ 25.7 & $>$ 25.4 &         25.5$\pm$0.3 & --- & 23.26$\pm$0.09   & 1.18$\pm$0.05 & $>$ 16.2 \\
SST24 J143001.9+334538 & $>$ 26.4 & $>$ 25.8 & $>$ 25.1 &             $>$ 26.5 & --- & 24.9$\pm$0.3         & 3.84$\pm$0.06 & $>$ 17.7 \\
SST24 J143025.7+342957 & 24.6$\pm$0.1 & 24.0$\pm$0.1 & 23.9$\pm$0.1 & 24.21$\pm$0.07 & --- & 22.29$\pm$0.03 & 2.47$\pm$0.05 &     15.4 \\
SST24 J143102.2+325152 & $>$ 25.7 & $>$ 25.2 & $>$ 25.2 &             $>$ 26.0 & --- & $>$ 25.1             & 1.19$\pm$0.05 & $>$ 15.8 \\
SST24 J143109.7+342802 & $>$ 26.4 & $>$ 25.5 & $>$ 25.2 &             $>$ 26.3 & --- & 23.6$\pm$0.1         & 1.11$\pm$0.04 & $>$ 16.0 \\
SST24 J143135.2+325456 & 24.7$\pm$0.1 & 23.9$\pm$0.1 & 23.5$\pm$0.1 & 25.1$\pm$0.2 & --- & 22.04$\pm$0.03   & 1.51$\pm$0.05 &     14.8 \\
SST24 J143225.3+334716 & $>$ 26.9 & $>$ 25.3 & $>$ 25.2 &             --- & 26.0$\pm$0.4 & 22.8$\pm$0.1     & 1.28$\pm$0.05 & $>$ 16.0 \\
SST24 J143242.5+342232 & $>$ 26.5 & $>$ 25.3 & $>$ 25.1 &             $>$ 26.7 & --- & 22.68$\pm$0.04       & 0.91$\pm$0.04 & $>$ 15.6 \\
SST24 J143251.8+333536 & $>$ 26.5 & $>$ 25.5 & $>$ 25.1 &             $>$ 26.5 & --- & 22.20$\pm$0.03       & 0.82$\pm$0.04 & $>$ 15.7 \\
SST24 J143312.7+342011 & 24.5$\pm$0.1 & 24.2$\pm$0.2 & 23.9$\pm$0.1 & 24.7$\pm$0.1 & --- & 22.23$\pm$0.04   & 1.76$\pm$0.04 &     15.2 \\
SST24 J143325.8+333736 & 25.6$\pm$0.3 & 24.6$\pm$0.3 & 23.9$\pm$0.2 & 25.6$\pm$0.3 & --- & 21.52$\pm$0.03   & 1.87$\pm$0.06 &     15.6 \\
SST24 J143358.0+332607 & $>$ 26.7 & $>$ 25.9 & $>$ 25.3 &             --- & $>$ 25.9 & 23.09$\pm$0.06       & 1.07$\pm$0.04 & $>$ 16.3 \\
SST24 J143447.7+330230 & $>$ 26.6 & $>$ 26.1 & $>$ 25.2 &             $>$ 26.0 & --- & 23.1$\pm$0.1         & 1.71$\pm$0.04 & $>$ 17.1 \\
SST24 J143504.1+354743 & $>$ 26.6 & $>$ 25.8 & $>$ 25.5 &             $>$ 26.4 & --- & 23.08$\pm$0.09       & 1.26$\pm$0.05 & $>$ 16.5 \\
SST24 J143508.4+334739 & 24.6$\pm$0.1 & 24.1$\pm$0.1 & 23.8$\pm$0.1 & 23.87$\pm$0.05 & --- & 22.69$\pm$0.07 & 2.65$\pm$0.08 &     15.6 \\
SST24 J143520.7+340418 & 24.8$\pm$0.1 & $>$ 25.1 & 24.1$\pm$0.2 &     $>$ 26.2 & --- & 24.0$\pm$0.2         & 1.53$\pm$0.06 & $>$ 15.9 \\
SST24 J143523.9+330706 & $>$ 26.8 & 25.0$\pm$0.2 & 24.9$\pm$0.3 &     --- & 24.7$\pm$0.2 & 22.93$\pm$0.06   & 1.09$\pm$0.05 &     15.5 \\
SST24 J143539.3+334159 & $>$ 26.4 & $>$ 25.5 & 24.7$\pm$0.3 &         25.1$\pm$0.2 & --- & 23.1$\pm$0.1     & 2.67$\pm$0.06 & $>$ 16.9 \\
SST24 J143545.1+342831 & $>$ 26.9 & $>$ 25.2 & $>$ 25.0 &             --- & $>$ 26.7 & 22.59$\pm$0.05       & 1.95$\pm$0.05 & $>$ 16.3 \\
SST24 J143644.2+350627 & 24.7$\pm$0.1 & 24.3$\pm$0.1 & 24.0$\pm$0.2 & 25.6$\pm$0.2 & --- & 22.70$\pm$0.07   & 2.34$\pm$0.05 &     15.6 \\
SST24 J143725.1+341502 & 25.4$\pm$0.2 & $>$ 25.4 & $>$ 25.2 &         --- & 26.1$\pm$0.6 & 22.70$\pm$0.08   & 1.41$\pm$0.05 & $>$ 16.2 \\
SST24 J143808.3+341016 & 25.1$\pm$0.1 & 24.4$\pm$0.3 & 23.9$\pm$0.1 & --- & 24.7$\pm$0.2 & 22.34$\pm$0.08   & 1.71$\pm$0.05 &     15.4 \\
\enddata
\tablenotetext{a}{ magnitude lower limits represent 2$\sigma$ values.}
\tablenotetext{b}{ magnitudes given in AB system.}
\tablenotetext{c}{ $R-[24]$ color in Vega system.}
\label{tab:sources}
\end{deluxetable*}

\subsection{Morphologies}

\subsubsection{GALFIT Results}\label{sec:galfit}

The results of our GALFIT analysis for the extended component S\'{e}rsic
profile fit to the NICMOS images are shown in Table~\ref{tab:galfit}, along
with 1-$\sigma$ uncertainties in the best-fit parameters.  The S\'{e}rsic
indices ($n$) range from 0.1 to 2.2 (median $n$ = 0.9).  For those objects
where $n < 1$, we note that constraining $n$ to be equal to 1 does not
significantly alter the remaining fit parameters.  As the S\'{e}rsic index
decreases, the radial profile flattens more rapidly within $r < R_{\rm eff}$
and the intensity drops more steeply beyond $r > R_{\rm eff}$.  For reference,
$n=0.5$ corresponds to a Gaussian profile, $n=1$ corresponds to an exponential
profile, and $n=4$ corresponds to a de Vaucouleurs profile
\citep{2002AJ....124..266P}.  The ratio of the minor to major axis ranges from
0.20 to 0.88 with a median value of 0.53.  In comparison, simulated merger
remnants tend to have a luminous component in the shape of an oblate spheroid,
with axis ratios of 1:1:0.5 \citep{2006ApJ...646L...9N}.  The projected axial
ratio should thus vary between 0.5 and 1.0.  Our observed median value of
$\approx$0.5 suggests that DOGs have more disk-like profiles than the simulated
merger remnants.  This may be due to a non-merger origin for DOGs, or it may be
an indication that DOGs have not progressed to the merger remnant stage.  

It is possible for a degeneracy to arise in the fitting parameters, in the
sense that a high-$n$, low point source fraction model may be comparable to a
low-$n$, high point source fraction model.  We have tested this by running
GALFIT with the point source fraction set to zero (i.e., removing the PSF
component).  The resulting $n$ values range from 0.1 to 5.2 with a median of
2.0, which is still below the value of 4 that is typical of early-type
galaxies.  A total of 7 DOGs have $n > 3$ using this zero-point-source model.
The best-fit $R_{\rm eff}$ values change by less than one pixel, with an offset
of -0.2$\pm$0.8 pixels.  In general, as mentioned in section~\ref{sec:parmeth},
the removal of the PSF component leads to larger reduced-$\chi^2$ values (see
columns 5 and 6 in Tab.~\ref{tab:galfit}).  However, we note that only one case
(SST24 J142644.3+333051) is associated with a $>$0.08 decrease in $\chi^2_\nu$
after adding a non-zero PSF component.  This suggests that the PSF component in
most of the DOGs in this sample is not dominant at rest-frame optical
wavelengths.

\begin{figure*}[!tbp]
\epsscale{1.20}
\plotone{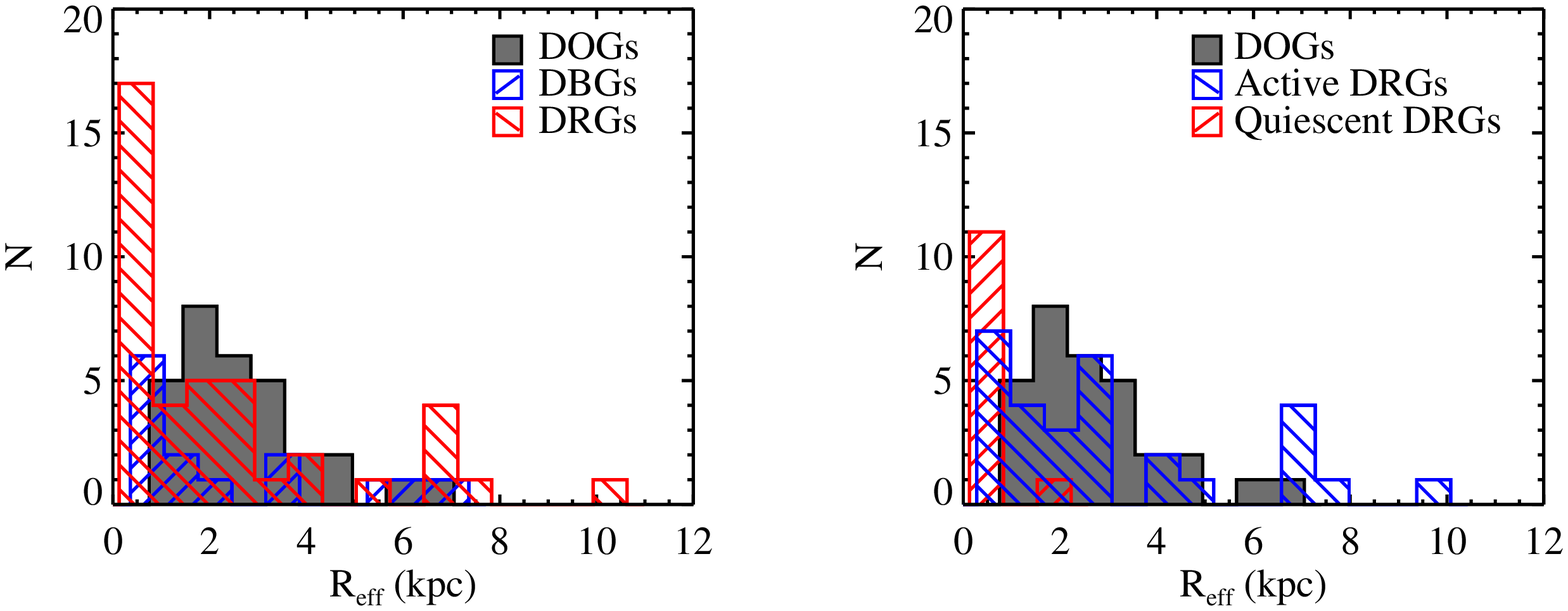}

\caption{ {\it Left:} Distribution of effective radius, $R_{\rm eff}$, for an
unconstrained S\'{e}rsic profile matched to the DOGs using GALFIT (filled grey
region), DBGs (Distant Blue Galaxies, diagonal hatched blue region) and DRGs
(opposite diagonal hatched red region).  {\it Right:} Distribution of $R_{\rm
eff}$ values for DOGs (filled grey region), active DRGs (diagonal hatched blue
region), and quiescent DRGs (opposite diagonal hatched red region).  DBG and
DRG data from \citet{2007ApJ...656...66Z} and \citet{2007ApJ...671..285T}.}

\label{fig:rpetmag}

\end{figure*}

We find the effective radius, $R_{\rm eff}$, ranges from 1.1 to 5.9~kpc with a
median value of 2.5~kpc.  In the left panel of Figure~\ref{fig:rpetmag}, we
show the distribution of $R_{\rm eff}$ values for DOGs (dark shaded histogram),
Distant Blue Galaxies \citep[diagonal blue hatched; DBGs, i.e., galaxies with
$z_{\rm phot} > 2$ that satisfy $J - Ks < 2.3$;][]{2007ApJ...671..285T}, and
DRGs \citep[opposite diagonal red hatched;][]{2007ApJ...656...66Z,
2007ApJ...671..285T}.  Two-sided K-S tests show that DOGs are dissimilar from
both populations, with a $<$7\% and $<$4\% chance of being drawn from the same
parent distribution, respectively.  Based on the nature of their UV-NIR SED,
DRGs may be separated into those that are actively forming stars (active DRGs
or sDRGs) and those that are not \citep[quiescent DRGs or qDRGs,
see][]{2007ApJ...656...66Z, 2007ApJ...671..285T}.  The right panel of
Figure~\ref{fig:rpetmag} shows the DOG $R_{\rm eff}$ distribution in comparison
to active DRGs (diagonal blue hatched) and quiescent DRGs (opposite diagonal
red hatched).  Quiescent DRGs have much smaller effective radii, while active
DRGs are much closer to DOGs.  Here the two-sided K-S test gives a 99.9994\%
and 34\% chance of being drawn from different parent distributions,
respectively, suggesting that the DOG and active DRG populations may overlap.

\begin{deluxetable*}{lcccccc}
\tabletypesize{\scriptsize} 
\tablecolumns{7}
\tablewidth{290pt}
\tablecaption{GALFIT Results}
\tablehead{
\colhead{} & \colhead{} & \colhead{} &
\colhead{$R_{\rm eff}$} \\
\colhead{} & \colhead{$n$} & \colhead{Axial Ratio} & \colhead{(kpc)} &
\colhead{$\chi^2_\nu$ \tablenotemark{a}} &
\colhead{$\chi^2_\nu$ \tablenotemark{b}} &
\colhead{$N_{\rm dof}$}
}
\startdata
SST24 J142538.2+351855 & 0.7$\pm$0.2 & 0.64$\pm$0.09 & 2.5$\pm$0.4 & 0.35 & 0.36 & 1653 \\
SST24 J142622.0+345249 & 0.1$\pm$0.1 & 0.55$\pm$0.04 & 2.5$\pm$1.0 & 0.92 & 0.92 & 1671 \\
SST24 J142626.4+344731 & 0.8$\pm$0.2 & 0.59$\pm$0.07 & 3.0$\pm$0.4 & 1.35 & 1.35 & 1671 \\
SST24 J142644.3+333051 & 5.6$\pm$3.4 & 0.71$\pm$0.07 & 1.1$\pm$0.3 & 1.16 & 2.58 & 1671 \\
SST24 J142645.7+351901 & 0.1$\pm$0.1 & 0.41$\pm$0.03 & 4.6$\pm$0.2 & 1.08 & 1.08 & 1671 \\
SST24 J142648.9+332927 & 0.7$\pm$0.3 & 0.70$\pm$0.05 & 1.8$\pm$0.1 & 1.08 & 1.08 & 1561 \\
SST24 J142653.2+330220 & 0.4$\pm$0.1 & 0.85$\pm$0.03 & 2.9$\pm$0.1 & 0.95 & 0.96 & 1671 \\
SST24 J142804.1+332135 & 1.0$\pm$0.6 & 0.19$\pm$0.06 & 6.4$\pm$2.6 & 0.95 & 0.96 & 1671 \\
SST24 J142924.8+353320 & 0.8$\pm$0.3 & 0.35$\pm$0.06 & 2.8$\pm$0.4 & 1.04 & 1.07 & 1671 \\
SST24 J142958.3+322615 & 0.4$\pm$0.1 & 0.78$\pm$0.03 & 2.1$\pm$0.1 & 0.36 & 0.39 & 1671 \\
SST24 J143001.9+334538 & 1.0$\pm$1.3 &   0.6$\pm$0.1 & 1.2$\pm$0.3 & 0.49 & 0.49 & 1671 \\
SST24 J143025.7+342957 & 1.0$\pm$0.1 & 0.52$\pm$0.02 & 1.8$\pm$0.1 & 0.90 & 0.96 & 1671 \\
SST24 J143102.2+325152 & ---       & ---         & ---       & ---  & ---  & ---  \\
SST24 J143109.7+342802 & 1.2$\pm$0.4 & 0.36$\pm$0.05 & 7.0$\pm$1.7 & 1.15 & 1.15 & 1671 \\
SST24 J143135.2+325456 & 0.5$\pm$0.1 & 0.53$\pm$0.01 & 3.8$\pm$0.1 & 1.54 & 1.56 & 1671 \\
SST24 J143225.3+334716 & 0.9$\pm$0.2 & 0.55$\pm$0.03 & 1.9$\pm$0.1 & 0.93 & 0.94 & 1671 \\
SST24 J143242.5+342232 & 2.0$\pm$0.4 & 0.50$\pm$0.06 & 4.0$\pm$0.8 & 0.57 & 0.58 & 1601 \\
SST24 J143251.8+333536 & 0.8$\pm$0.1 & 0.50$\pm$0.02 & 3.3$\pm$0.1 & 0.50 & 0.52 & 1671 \\
SST24 J143312.7+342011 & 0.7$\pm$0.1 & 0.51$\pm$0.01 & 4.0$\pm$0.1 & 0.63 & 0.66 & 1671 \\
SST24 J143325.8+333736 & 1.3$\pm$0.1 & 0.70$\pm$0.01 & 3.1$\pm$0.1 & 0.52 & 0.60 & 1671 \\
SST24 J143358.0+332607 & 2.1$\pm$0.5 & 0.84$\pm$0.07 & 1.4$\pm$0.1 & 0.49 & 0.48 & 1671 \\
SST24 J143447.7+330230 & 0.4$\pm$0.1 & 0.88$\pm$0.04 & 1.9$\pm$0.1 & 0.88 & 0.89 & 1671 \\
SST24 J143504.1+354743 & 0.5$\pm$0.1 & 0.43$\pm$0.03 & 4.6$\pm$0.3 & 0.91 & 0.93 & 1671 \\
SST24 J143508.4+334739 & 2.6$\pm$0.4 & 0.40$\pm$0.03 & 2.2$\pm$0.2 & 1.06 & 1.10 & 1671 \\
SST24 J143520.7+340418 & 0.6$\pm$0.4 &   0.5$\pm$0.1 & 3.4$\pm$0.7 & 1.08 & 1.10 & 1671 \\
SST24 J143523.9+330706 & 0.5$\pm$0.1 & 0.46$\pm$0.02 & 1.5$\pm$0.1 & 1.00 & 1.05 & 1671 \\
SST24 J143539.3+334159 & 1.8$\pm$0.5 & 0.82$\pm$0.09 & 3.7$\pm$0.7 & 0.91 & 0.92 & 1671 \\
SST24 J143545.1+342831 & 1.1$\pm$0.2 & 0.49$\pm$0.03 & 2.2$\pm$0.1 & 1.04 & 1.06 & 1671 \\
SST24 J143644.2+350627 & 0.9$\pm$0.2 & 0.70$\pm$0.05 & 1.9$\pm$0.1 & 1.22 & 1.23 & 1670 \\
SST24 J143725.1+341502 & 0.3$\pm$0.1 & 0.50$\pm$0.03 & 3.4$\pm$0.2 & 1.09 & 1.14 & 1424 \\
SST24 J143808.3+341016 & 0.9$\pm$0.1 & 0.81$\pm$0.04 & 3.0$\pm$0.2 & 1.37 & 1.38 & 1671 \\
\enddata
\tablenotetext{a}{Asssuming finite PSF contribution}
\tablenotetext{b}{Asssuming zero PSF contribution}
\label{tab:galfit}
\end{deluxetable*}

In Table~\ref{tab:psf}, we give the $V$, $I$, and $H$ magnitudes of the nuclear
(PSF) component and the extended (galaxy) component, as well as the fraction of
light contributed by a point source (including 1-$\sigma$ uncertainties).  When
the nuclear component is not detected, we quote the 3-$\sigma$ limit on the
point source fraction.  The magnitude of the PSF component is measured using
the same aperture photometry method described in Section~\ref{sec:photo}, with
the exception that the sky background is assumed to be zero.  For the extended
component, we subtract the PSF component from the science image and compute the
photometry in the usual way on the residual image.
The fraction of light due to an unresolved component in the rest-optical ranges
from 0.04 to 0.78, and the median is 0.12.  In the rest-UV, however, this
fraction is signicantly smaller, with only one object having a detected
fraction.  This object, SST24 J143523.9+330706, stands out as unique by virtue of
having a greater point source fraction in the rest-UV than in the rest-optical.
This behavior is unique within our sample (but is expected when the AGN is
viewed without obscuration) and is also reflected in the non-parametric
measures of its morphology (see Sect.~\ref{sec:nonpar} for more detail).

\begin{deluxetable*}{lcccccccc}
\tabletypesize{\scriptsize} 
\tablecolumns{9}
\tablewidth{460pt}
\tablecaption{PSF Subtraction Analysis}
\tablehead{
\colhead{Source Name} & 
\colhead{$V_{\rm nuc}$} & 
\colhead{$I_{\rm nuc}$} & 
\colhead{$H_{\rm nuc}$} & 
\colhead{$V_{\rm gal}$} & 
\colhead{$I_{\rm gal}$} & 
\colhead{$H_{\rm gal}$} & 
\colhead{$f_{\rm PSF}^{\rm opt}$} & 
\colhead{$f_{\rm PSF}^{\rm ir}$}
}
\startdata
SST24 J142538.2+351855 &       $>$ 28.3 & ---      & 27$\pm$2 &       $>$ 26.0 & ---      & 24.1$\pm$0.3 & --- & $<$0.27 \\
SST24 J142622.0+345249 & ---      &       $>$ 27.9 & 27$\pm$1 & ---      & 24.2$\pm$0.1 & 23.6$\pm$0.2 & $<$0.01 & $<$0.15 \\
SST24 J142626.4+344731 &       $>$ 27.6 & ---      & 26.8$\pm$0.3 &       $>$ 26.4 & ---      & 23.7$\pm$0.2 & --- & $<$0.07 \\
SST24 J142644.3+333051 &       $>$ 28.1 & ---      & 22.3$\pm$0.1 & 26.1$\pm$0.6 & ---      & 23.4$\pm$0.1 & $<$0.06 & 0.73$\pm$0.07 \\
SST24 J142645.7+351901 &       $>$ 27.7 & ---      & 27$\pm$1 &       $>$ 26.2 & ---      & 22.9$\pm$0.1 & --- & $<$0.08 \\
SST24 J142648.9+332927 & ---      &       $>$ 28.1 & 25.2$\pm$0.3 & ---      & 24.9$\pm$0.4 & 23.3$\pm$0.2 & $<$0.03 & 0.15$\pm$0.06 \\
SST24 J142653.2+330220 & ---      &       $>$ 28.3 & 25.8$\pm$0.2 & ---      & 25.0$\pm$0.4 & 22.8$\pm$0.2 & $<$0.01 & 0.06$\pm$0.03 \\
SST24 J142804.1+332135 &       $>$ 27.7 & ---      & 26.1$\pm$0.5 &       $>$ 26.6 & ---      &       $>$ 25.2 & --- & --- \\
SST24 J142924.8+353320 &       $>$ 27.9 & ---      & 25.9$\pm$0.5 &       $>$ 26.1 & ---      &       $>$ 25.3 & --- & --- \\
SST24 J142958.3+322615 &       $>$ 28.2 & ---      & 25.0$\pm$0.1 & 25.6$\pm$0.6 & ---      & 23.5$\pm$0.2 & $<$0.01 & 0.20$\pm$0.03 \\
SST24 J143001.9+334538 &       $>$ 27.9 & ---      & 26.6$\pm$0.9 &       $>$ 26.5 & ---      &       $>$ 25.5 & --- & $<$0.53 \\
SST24 J143025.7+342957 &       $>$ 28.4 & ---      & 23.9$\pm$0.1 & 24.2$\pm$0.2 & ---      & 22.58$\pm$0.07 & $<$0.01 & 0.24$\pm$0.03 \\
SST24 J143102.2+325152 &       $>$ 27.8 & ---      &       $>$ 27.0 &       $>$ 26.0 & ---      &       $>$ 25.1 & --- & --- \\
SST24 J143109.7+342802 &       $>$ 27.9 & ---      & 27$\pm$1 &       $>$ 26.3 & ---      & 23.78$\pm$0.3 & --- & $<$0.18 \\
SST24 J143135.2+325456 &       $>$ 27.8 & ---      & 26.1$\pm$0.3 & 25.2$\pm$0.5 & ---      & 22.1$\pm$0.1 & $<$0.06 & $<$0.02 \\
SST24 J143225.3+334716 & ---      &       $>$ 28.3 & 24.8$\pm$0.1 & ---      & 26$\pm$1 & 23.0$\pm$0.2 & $<$0.02 & 0.17$\pm$0.04 \\
SST24 J143242.5+342232 &       $>$ 28.5 & ---      & 25.4$\pm$0.2 &       $>$ 26.7 & ---      & 22.9$\pm$0.1 & --- & 0.09$\pm$0.03 \\
SST24 J143251.8+333536 &       $>$ 28.7 & ---      & 24.9$\pm$0.1 &       $>$ 26.5 & ---      & 22.4$\pm$0.1 & --- & 0.09$\pm$0.01 \\
SST24 J143312.7+342011 &       $>$ 27.9 & ---      & 24.6$\pm$0.1 & 24.7$\pm$0.3 & ---      & 22.4$\pm$0.1 & $<$0.04 & 0.12$\pm$0.02 \\
SST24 J143325.8+333736 &       $>$ 27.6 & ---      & 23.6$\pm$0.1 & 25.8$\pm$0.7 & ---      & 21.6$\pm$0.1 & --- & 0.13$\pm$0.02 \\
SST24 J143358.0+332607 & ---      & 27.5$\pm$0.5 & 25.8$\pm$0.3 & ---      &       $>$ 25.9 & 23.5$\pm$0.2 & --- & 0.10$\pm$0.05 \\
SST24 J143447.7+330230 &       $>$ 27.9 & ---      & 25.8$\pm$0.6 &       $>$ 26.0 & ---      & 23.2$\pm$0.2 & --- & $<$0.12 \\
SST24 J143504.1+354743 &       $>$ 27.7 & ---      & 24.9$\pm$0.2 &       $>$ 26.4 & ---      & 23.3$\pm$0.2 & --- & 0.18$\pm$0.05 \\
SST24 J143508.4+334739 & 27$\pm$2 & ---      & 24.9$\pm$0.3 & 23.9$\pm$0.1 & ---      & 22.6$\pm$0.1 & $<$0.14 & 0.11$\pm$0.04 \\
SST24 J143520.7+340418 &       $>$ 28.1 & ---      & 24.9$\pm$0.1 &       $>$ 26.2 & ---      & 25.0$\pm$0.5 & --- & 0.5$\pm$0.1 \\
SST24 J143523.9+330706 & ---      & 26.2$\pm$0.2 & 24.9$\pm$0.2 & ---      & 25.0$\pm$0.5 & 23.1$\pm$0.1 & 0.24$\pm$0.04 & 0.17$\pm$0.03 \\
SST24 J143539.3+334159 &       $>$ 28.4 & ---      & 25.2$\pm$0.2 & 25.2$\pm$0.5 & ---      & 23.3$\pm$0.3 & $<$0.01 & 0.15$\pm$0.06 \\
SST24 J143545.1+342831 & ---      &       $>$ 28.0 & 24.2$\pm$0.1 & ---      &       $>$ 26.7 & 22.9$\pm$0.1 & --- & 0.22$\pm$0.03 \\
SST24 J143644.2+350627 & 28$\pm$1 & ---      & 24.9$\pm$0.3 & 25.7$\pm$0.4 & ---      & 22.9$\pm$0.1 & $<$0.11 & 0.13$\pm$0.04 \\
SST24 J143725.1+341502 & ---      &       $>$ 28.1 & 23.6$\pm$0.1 & ---      &       $>$ 26.2 & 23.3$\pm$0.2 & --- & 0.43$\pm$0.05 \\
SST24 J143808.3+341016 & ---      &       $>$ 28.3 & 25.9$\pm$0.4 & ---      & 24.7$\pm$0.4 & 22.4$\pm$0.2 & $<$0.01 & $<$0.06 \\
\enddata
\label{tab:psf}
\end{deluxetable*}

\begin{figure*}[!tbp]
\epsscale{1.2}
\plotone{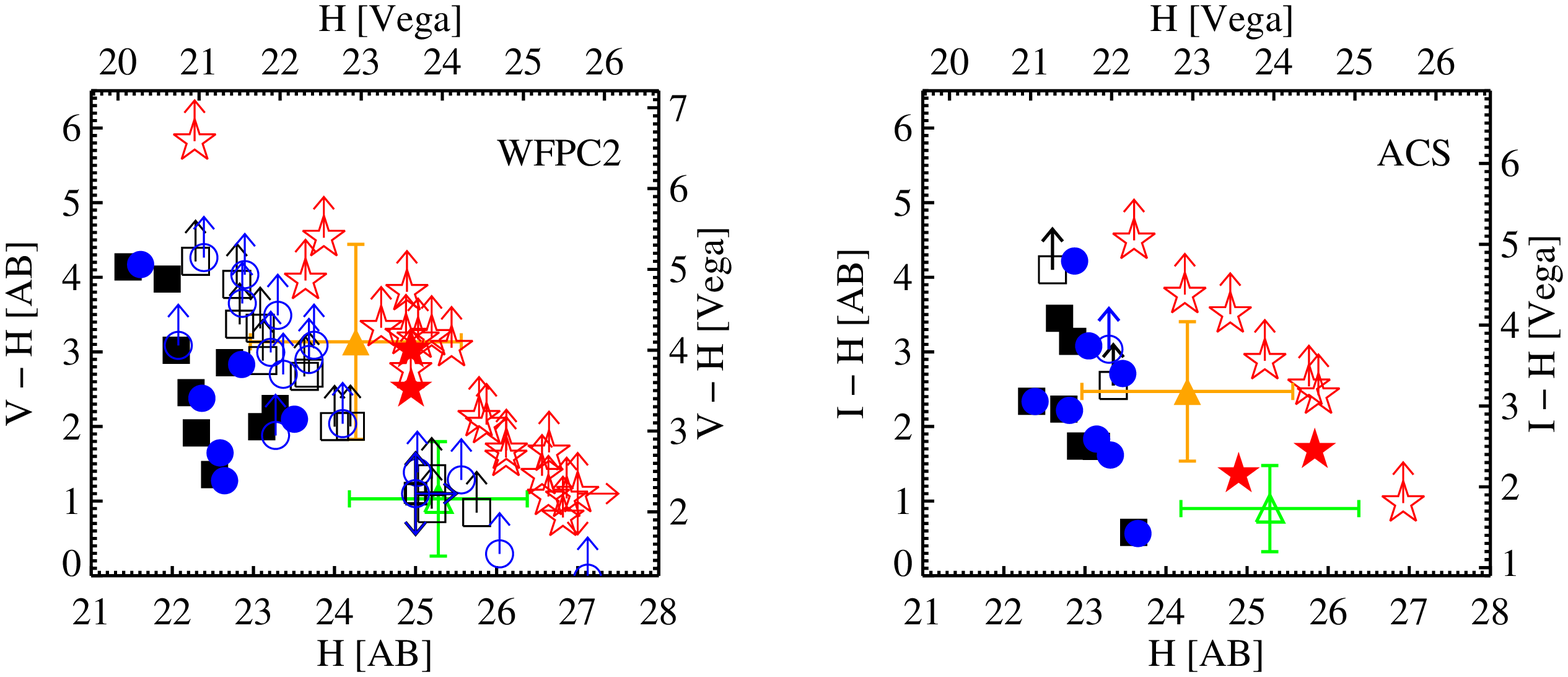}

\caption{Color-magnitude diagram for DOGs, broken down into an extended
component (unconstrained S\'ersic profile) and an unresolved nuclear component
(from TinyTim PSF).  {\it Left:} $V - H$ vs. $H$ for DOGs observed by WFPC2
(black squares).  The extended component of each DOG is shown with a blue
circle and the point source with a red star.  Detections (Lower limits) are
plotted with filled (open) symbols.  Open green triangles show the median and
1$\sigma$ dispersion in colors of high-$z$ galaxies from the HDF-N (Papovich,
personal communication and HDF-S \citep{2003AJ....125.1107L}.  The subset of
these galaxies qualifying as DRGs are shown with a filled orange triangle.
{\it Right:} $I - H$ vs.  $H$ for DOGs observed by ACS.  Symbols are same as in
left panel.  }

\label{fig:psf}

\end{figure*}

Figure~\ref{fig:psf} shows the $V-H$ and $I-H$ colors of the nuclear, extended,
and full galaxy components as a function of $H$, in AB magnitudes.  High-$z$
galaxies and DRGs in the HDF-N and HDF-S are also shown.  The full galaxy and
extended components have similar colors and $H$ magnitudes, consistent with the
nuclear component not dominating the flux.  This is why even the extended
components of DOGs are redder in both $V - H$ and $I-H$ compared to Lyman break
galaxies (LBGs) in the HDF, and suggests that one cannot create a DOG simply by
adding an obscured AGN to a star-forming galaxy like an LBG.  DRGs show greater
overlap with the colors of DOGs but few are as bright in $H$ as the DOGs in our
sample.

\subsubsection{Non-parametric Classification Results} \label{sec:nonpar}

Non-parametric methods of characterizing galaxy morphology are known to require
high S/N imaging to yield reliable results \citep{2004AJ....128..163L}.  In the
rest-UV, where the DOGs are very faint, none of the 22 WFPC2 images and only 6
out of 9 ACS images have the per-pixel S/N necessary to compute $R_p$, $G$,
$M_{20}$, and $C$.  In the rest-optical, however, DOGs are much brighter and 23
out of 31 NICMOS images have sufficient S/N.  Table~\ref{tab:morph} presents
the visual and non-parametric measures of DOG morphologies, including the
per-pixel S/N, $R_p$, $G$, $M_{20}$, and $C$ values for the ACS/WFPC2 and
NICMOS images.

\begin{deluxetable*}{lcccccccccccc}
\tabletypesize{\scriptsize} 
\tablecolumns{13}
\tablewidth{470pt}
\tablecaption{Non-parametric Morphological Classifications}
\tablehead{
\colhead{} & \multicolumn{6}{c}{ACS/WFPC2} & \multicolumn{6}{c}{NICMOS} \\
\colhead{Source Name} & 
\colhead{Visual\tablenotemark{a}} & 
\colhead{$S/N$} &
\colhead{$R_p$ ($\arcsec$)} & 
\colhead{$G$} & 
\colhead{$M_{20}$} & 
\colhead{$C$} & 
\colhead{Visual\tablenotemark{a}} & 
\colhead{$S/N$} &
\colhead{$R_p$ ($\arcsec$)} & 
\colhead{$G$} & 
\colhead{$M_{20}$} & 
\colhead{$C$}
}
\startdata
SST24 J142538.2+351855 &   TFTT       &  --- & --- & --- & --- & --- &    {\it Reg}   & --- & --- & --- & --- & --- \\
SST24 J142622.0+345249 &   Irr        &  5.2  & 0.6  & 0.45 & -0.8 & 3.6  & {\it Irr} & 3.0  & 0.5  & 0.44 & -1.0 & 2.1  \\
SST24 J142626.4+344731 &   Irr        &  --- & --- & --- & --- & --- &   {\it  Reg}   & --- & --- & --- & --- & --- \\
SST24 J142644.3+333051 &   TFTT       &  --- & --- & --- & --- & --- &     Reg        & 10.7 & 0.3  & 0.56 & -1.8 & 3.3  \\
SST24 J142645.7+351901 &   TFTT       &  --- & --- & --- & --- & --- &      Irr       & 2.7  & 0.8  & 0.42 & -0.7 & 4.3  \\
SST24 J142648.9+332927 &  {\it Irr}   &  2.7  & 0.6  & 0.46 & -0.8 & 4.9  & {\it Reg} & 2.6  & 0.6  & 0.50 & -1.7 & 2.9  \\
SST24 J142653.2+330220 &   Irr        &  2.2  & 0.8  & 0.42 & -0.7 & 3.6  & {\it Reg} & 3.9  & 0.6  & 0.43 & -1.2 & 2.2  \\
SST24 J142804.1+332135 &   TFTT       &  --- & --- & --- & --- & --- &      TFTT      & --- & --- & --- & --- & --- \\
SST24 J142924.8+353320 &   TFTT       &  --- & --- & --- & --- & --- &      Irr       & --- & --- & --- & --- & --- \\
SST24 J142958.3+322615 &  Reg         &  --- & --- & --- & --- & --- &     Reg        & 3.8  & 0.5  & 0.46 & -1.4 & 2.3  \\
SST24 J143001.9+334538 &   TFTT       &  --- & --- & --- & --- & --- &      TFTT      & --- & --- & --- & --- & --- \\
SST24 J143025.7+342957 &  Reg         &  --- & --- & --- & --- & --- &     Reg        & 6.1  & 0.6  & 0.54 & -1.7 & 2.9  \\
SST24 J143102.2+325152 &   TFTT       &  --- & --- & --- & --- & --- &      TFTT      & --- & --- & --- & --- & --- \\
SST24 J143109.7+342802 &   TFTT       &  --- & --- & --- & --- & --- &      Irr       & --- & --- & --- & --- & --- \\
SST24 J143135.2+325456 &   TFTT       &  --- & --- & --- & --- & --- &     {\it Irr}  & 2.8  & 1.3  & 0.52 & -2.5 & 4.7  \\
SST24 J143225.3+334716 &  {\it Irr}   &  3.9  & 0.4  & 0.37 & -1.6 & 2.4  &     Reg   & 5.1  & 0.5  & 0.50 & -1.4 & 2.8  \\
SST24 J143242.5+342232 &   TFTT       &  --- & --- & --- & --- & --- &      Irr       & --- & --- & --- & --- & --- \\
SST24 J143251.8+333536 &   TFTT       &  --- & --- & --- & --- & --- &    {\it Reg}   & 4.7  & 0.9  & 0.47 & -1.7 & 2.7  \\
SST24 J143312.7+342011 &   Irr        &  --- & --- & --- & --- & --- &    {\it Reg}   & 3.8  & 1.1  & 0.51 & -1.4 & 3.3  \\
SST24 J143325.8+333736 &  Reg         &  --- & --- & --- & --- & --- &     Reg        & 5.1  & 1.0  & 0.54 & -1.9 & -2.1\tablenotemark{b} \\
SST24 J143358.0+332607 &  Reg         &  --- & --- & --- & --- & --- &     Reg        & 4.1  & 0.5  & 0.50 & -1.4 & 3.1  \\
SST24 J143447.7+330230 &   TFTT       &  --- & --- & --- & --- & --- &    {\it Reg}   & 4.3  & 0.5  & 0.46 & -1.2 & 2.0  \\
SST24 J143504.1+354743 &   TFTT       &  --- & --- & --- & --- & --- &    {\it Reg}   & 2.2  & 0.9  & 0.49 & -0.9 & -2.1\tablenotemark{b} \\
SST24 J143508.4+334739 &  {\it  Irr}  &  --- & --- & --- & --- & --- &    {\it Reg}   & 4.1  & 0.8  & 0.53 & -1.2 & 2.9  \\
SST24 J143520.7+340418 &  Reg         &  --- & --- & --- & --- & --- &    {\it Reg}   & 3.3  & 0.4  & 0.47 & -0.9 & 3.3  \\
SST24 J143523.9+330706 & {\it Reg}    &  5.2  & 0.5  & 0.56 & -1.2 & 2.8  & {\it Irr} & 6.3  & 0.4  & 0.47 & -1.2 & 2.3  \\
SST24 J143539.3+334159 &   TFTT       &  --- & --- & --- & --- & --- &    Reg         & 2.7  & 0.7  & 0.50 & -1.7 & 3.4  \\
SST24 J143545.1+342831 &   TFTT       &  --- & --- & --- & --- & --- &    {\it Reg}   & 4.6  & 0.6  & 0.55 & -1.6 & 3.2  \\
SST24 J143644.2+350627 &   TFTT       &  --- & --- & --- & --- & --- &    {\it Reg}   & 3.1  & 0.6  & 0.52 & -1.7 & 2.7  \\
SST24 J143725.1+341502 &   TFTT       &  --- & --- & --- & --- & --- &     Reg        & 2.5  & 0.9  & 0.52 & -2.1 & 3.7  \\
SST24 J143808.3+341016 &   Irr        &  2.3  & 1.3  & 0.44 & -0.9 & 3.6  & Irr       & 2.3  & 1.0  & 0.48 & -1.6 & 3.2  \\

\enddata
\tablenotetext{a}{ Mode of visual classification.  Italics indicate
multiple users disagreed with the mode.}
\tablenotetext{b}{ Negative $C$ value indicates $r_{20}$ was too small to be measured accurately.}
\label{tab:morph}
\end{deluxetable*}

%


In the left panel of Figure~\ref{fig:gm20}, we plot $G$ as a function of
$M_{20}$ as measured in the rest-UV for DOGs, a field galaxy sample, and
simulated $r^{1/4}$ bulges and pure exponential disks
{\citep{2006ApJ...636..592L}.  None of the DOGs fall within the pure
exponential disk- or $r^{1/4}$ bulge-dominated regime.  The field galaxy
population is composed of sources identified within the ACS FOVs and is
selected to span the same magnitude range as the DOGs in our sample.   We use
our NDWFS data to apply color cuts in $B_W-R$ and $R-I$ space in order to
remove objects with colors typical of $z < 0.7$ sources.  The morphologies of
this field galaxy sample are represented with gray contours.  Four out of six
DOGs lie in the lower left corner of the plot, with low $G$ and high $M_{20}$
values indicating irregular, diffuse morphologies.  In general, low $G$ and
high $M_{20}$ values are indicative of dust-enshrouded stellar populations,
where obscuration by dust causes a galaxy to appear very clumpy with flux
distributed among many pixels \citep{2008arXiv0805.1246L}.  

\begin{figure*}[!tbp]
\epsscale{1.20}
\plotone{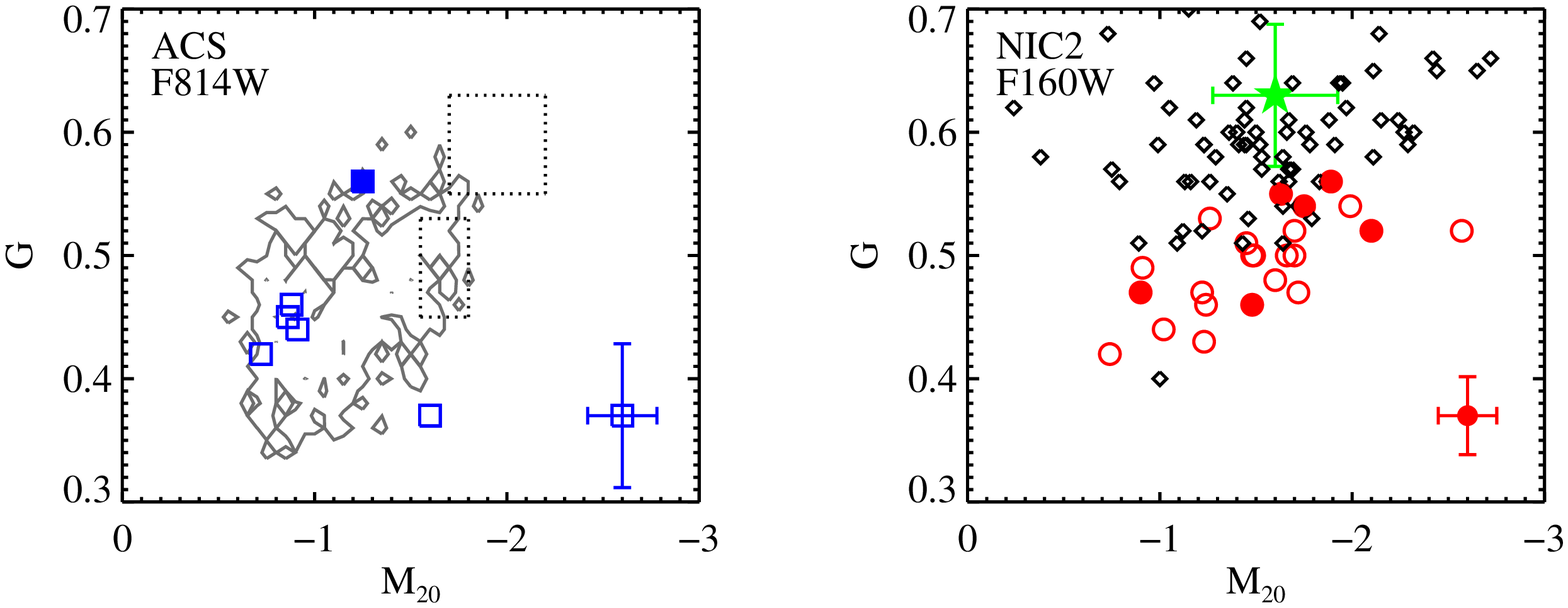}

\caption{Gini coefficient vs. $M_{20}$.  {\it Left:} Morphological measures
from ACS/F814W images of DOGs (blue squares) and field galaxies (grey solid
contours, see text).  Representative error bars in the lower right corner
include uncertainties due both to low S/N and low spatial resolution and are
estimated using a method similar to that in \citet{2004AJ....128..163L}.
Filled symbols have greater than 20\% point-source contribution.  Top and
bottom dotted boxes show where simulated face-on bulges and disks lie,
respectively \citep{2006ApJ...636..592L}.  {\it Right:} Same plot but showing
results from NIC2/F160W images of DOGs (red circles), HDF-N LBGs (filled green
star) and a sample of local ULIRGs (black diamonds)
\citep{2004AJ....128..163L}.}

\label{fig:gm20}

\end{figure*}

One object (SST24 J143523.9+330706, panel 27 in Fig.~\ref{fig:cutouts1}) has a
higher $G$ and lower $M_{20}$ value than nearly all of the field galaxies.
This is the same object that shows a stronger point-source contribution in the
rest-UV than the rest-optical.  Visual inspection of this object's cutout image
reveals an extended feature fading towards the southeast that is present in
both the rest-UV and rest-optical.  It appears that the central activity in
this source is not quite as obscured as in other DOGs, but it is not yet clear
why this is the case.

The right panel in Figure~\ref{fig:gm20} shows $G$ and $M_{20}$ values as
measured in the rest-optical for DOGs as well as LBGs in the HDF-N and a sample
of local ($z < 0.1$) ULIRGs \citep{2004AJ....128..163L}.  DOGs shift to more
typical morphological parameters in the rest-optical compared to the rest-UV,
but they are offset from the parameter space occupied by LBGs and local ULIRGs.
The median $G$ and $M_{20}$ values for DOGs are 0.49 and $-$1.24, respectively,
while for LBGs they are 0.63 and $-$1.6 and for local ULIRGs they are 0.59 and
$-$1.5.  Part of the difference in $G$ and $M_{20}$ compared to LBGs may be
that LBGs are more compact, and hence less resolved.  The offset to lower $G$
values in DOGs compared to local ULIRGs is remarkable and indicates that either
different mechanisms are involved in creating these two populations, or they
represent different stages in the evolution of massive galaxies.  We note that
the ULIRG sample has comparable S/N as the DOGs studied here, and that the
rest-frame wavelength of both samples is similar ($\sim$7000~\AA~vs.
$\sim$5300~\AA, respectively).  While there is a greater relative difference
in the spatial resolution of the two samples ($\approx$0.2~kpc~pix$^{-1}$ vs.
($\approx$0.6~kpc~pix$^{-1}$, respectively), \citet{2004AJ....128..163L} found
that systematic offsets at these resolutions should be on the order of 20\% or
less for $C$ and $M_{20}$ and less than 10\% for $G$.  Therefore, the
difference in $G$ cannot be explained by spatial resolution effects alone.

As a qualitative consistency check, we examined $R$-band images of a sample of
56 ULIRGs from \citet{1996AJ....111.1025M} and determined that 20 (35\%) have
double nuclei with nuclear separations larger than 2.3~kpc, approximately the
resolution limit of our NIC2 images.  In comparison, only one DOG has two
well-detected, distinct nuclei and three or four have low S/N components
separated by 0$\farcs$5 ($\approx$4~kpc), implying that at most 16\% of the DOGs
in our sample have multiple nuclei with separations larger than 2.3~kpc.  This
result is qualitatively consistent with the differences seen in $G$ between
local ULIRGs and DOGs.  An important caveat with this analysis is that our
sample of DOGs is dominated by power-law sources, while the ULIRG sample has a
variety of rest-frame NIR SED shapes.  For reference, we measured the $G$ and
$M_{20}$ values of a well-detected, non-DOG point source (S/N per pixel of 18)
in one of our NIC2 images, and found values of 0.62 and $-$1.7, respectively.
The DOG whose morphology is dominated by a point source (see panel 31 in
Fig.~\ref{fig:cutouts1}) has a lower $G$ value (0.56) but almost the same
$M_{20}$ ($-$1.8).  This may be an indication that this DOG contains an
underlying extended component, but the data are not conclusive.

\section{Discussion}\label{sec:disc}

\subsection{Dust and Stellar Mass Estimates}\label{sec:masses}

Here we estimate some of the intrinsic properties of DOGs, including lower
limits on their reddening ($A_V$), dust and gas masses, and stellar masses.  To
do this, we use Simple Stellar Population (SSP) template SEDs from the
\citet{2003MNRAS.344.1000B} population synthesis library with ages spaced
logarithmically from 10~Myr up to 1~Gyr, as well as the median QSO template from
\citet{1994ApJS...95....1E}.   All models used here have solar metallicity, a
Chabrier IMF over the mass range 0.1-100$M_\sun$ \citep{2003PASP..115..763C},
and use the Padova~1994 evolutionary tracks \citep{1996A&AS..117..113G}.  The
reddening law used is a combination of that from \citet{2000ApJ...533..682C} and
longer wavelength estimates from \citet{2003ARA&A..41..241D}, and assumes the
case of a dust screen in front of the emitting source in order to derive a firm
lower limit on $A_V$.

For each DOG, we estimate $A_V$ needed as a function of age by determining the
amount of extinction necessary to redden the given SSP template such that it
reproduces the observed $V-H$ or $I-H$ color.  The process is illustrated in
Figure~\ref{fig:redden}.  Each panel shows the colors of DOGs in our sample as a
function of redshift.  Blue circles represent the extended
component and red stars show the point source component of each DOG, as
described in section~\ref{sec:galfit}.  Dotted lines show the expected colors
of the SSP templates for varying amounts of extinction.  Even with no
extinction ($A_V = 0$), the oldest SSP templates are too red to reproduce the
colors exhibited by DOGs.  The QSO templates, on the other hand, require large
$A_V$ values in order to match the DOG nuclear colors.

\begin{figure*}[!tbp]
\epsscale{1.00}
\plotone{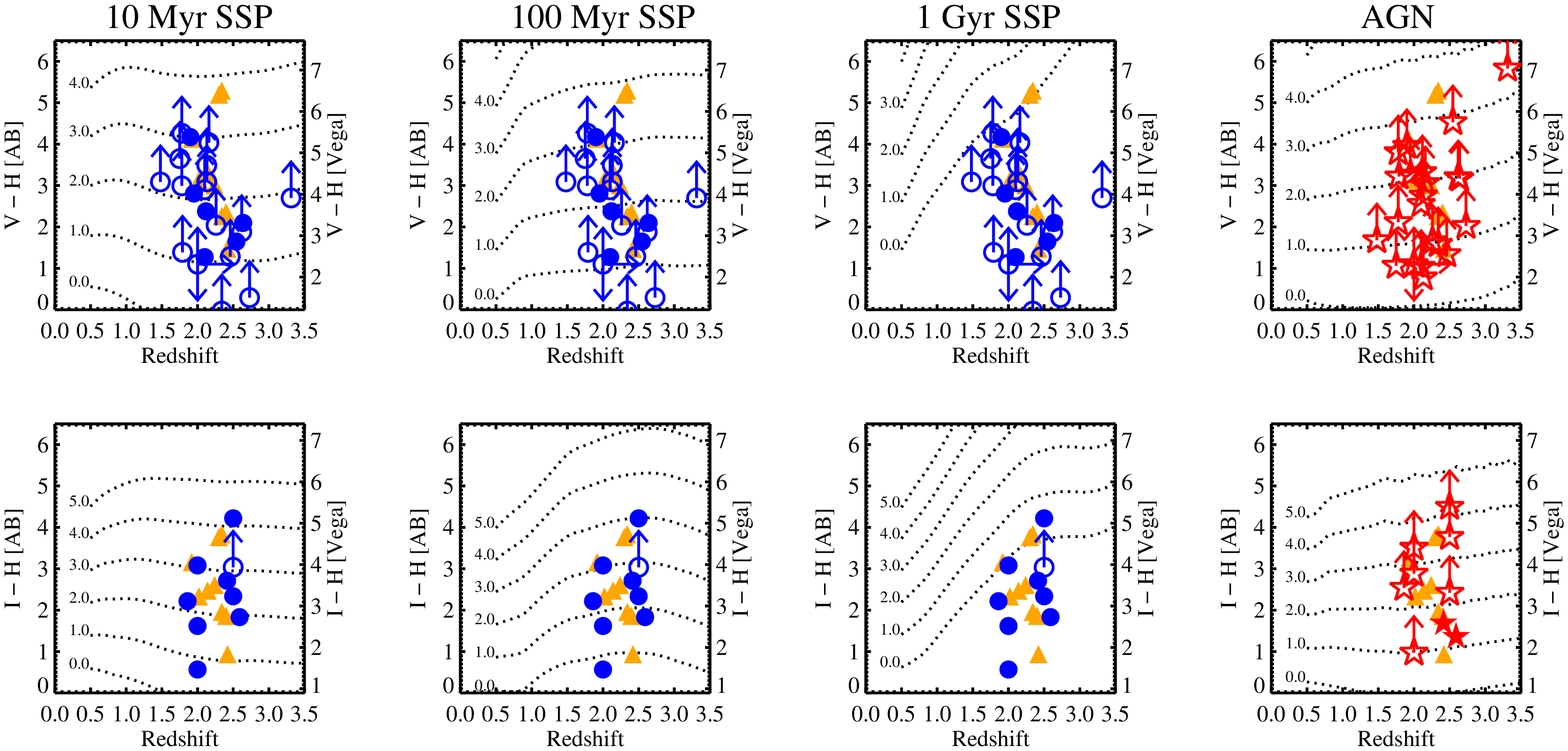}

\caption{$V-H$ (top row) and $I-H$ (bottom row) as a function of spectroscopic
redshift for each DOG.  The first three columns show the colors of the extended
component, while the fourth column shows the colors of the unresolved component
(filled symbols are detections, open symbols are lower limits).  Filled orange
triangles represent DRGs in the HDF-S.  Dotted lines trace the evolution of
colors with redshift of reddened simple stellar population models from
\citet{2003MNRAS.344.1000B} with solar metallicty, a Chabrier IMF and at ages
of 10~Myr, 100~Myr, and 1~Gyr (three columns on left), as well as of the median
QSO template from \citet{1994ApJS...95....1E}.   }

\label{fig:redden}

\end{figure*}

We use the relation from \citet{1978ApJ...224..132B} to convert $A_V$ to the
total column density of hydrogen atoms and molecules, $N_{\rm H}$.  For the
100~Myr SSP, the column densities range from $3 \times 10^{20} - 6 \times
10^{21} \:$cm$^{-2}$, with a median $N_{\rm H}$ of $3 \times 10^{21}
\:$cm$^{-2}$.  If we assume the dust is distributed in a spherical shell
around the source with radius equal to the effective radius, then we can
place a lower limit on the dust mass:

\begin{equation}
M_{\rm dust} \geq \frac{1}{f_{\rm gd}} \mu_p N_{\rm H} \times 4\pi R_{\rm
eff}^2.
\label{eq:mdust_av}
\end{equation}

\noindent Here, $f_{\rm gd}$ is the gas-to-dust mass ratio and $\mu_p$ is the
mean molecular weight of the gas, which we take to be 1.6$m_p$, where $m_p$ is
the mass of a proton.  We adopt the average gas-to-dust mass ratio of 120
measured for the nuclear regions of local ULIRGs by
\citet{2008arXiv0806.3002W}.  We find dust mass lower limits ranging from $2
\times 10^5 - 6 \times 10^7 M_\sun$, with a median of $9 \times 10^6 \: M_\sun$
for a 100~Myr SSP.

A complementary method of estimating the dust mass is based on measurements of
optically thin sub-mm emission.  Because sub-mm photometry for the DOGs in our
sample is currently unavailable, we extrapolate from the 24$\mu$m flux density
measurement.  We used the Mrk231 template to determine the extrapolated
850$\mu$m flux density, $F_{850}$.  Using a template SED of a galaxy with
colder dust such as Arp220 would increase the inferred 850$\mu$m flux.  We
follow \citet{1997MNRAS.289..766H} and estimate the dust mass using:

\begin{equation}
M_{\rm dust} = \frac{1}{1+z} \frac{F_{850} d_L^2} { \kappa_d B(\nu,T_{\rm d}) },
\label{eq:mdust_24}
\end{equation}

\noindent where $d_L$ is the luminosity distance, $\kappa_d$ is the
rest-frequency mass absorption coefficient, and $B(\nu,T)$ is the value of the
modified blackbody function ($\beta =1.5$) at the rest frequency $\nu$ and a
temperature $T$.  The appropriate $\kappa_d$ value is interpolated from
\citet{2003ARA&A..41..241D}, with typical values being 5~cm$^{2}$~g$^{-1}$.
There is at least a factor of 2 uncertainty in this quantity.  We have assumed
relatively hot dust ($T_{\rm d} = 75$~K), since we expect AGN heating to play an
important role in this sample of DOGs (a dust temperature of 50~K would increase
the inferred dust mass by an additional factor of $\approx$1.5).  Using this
method, we find dust masses of $8\times10^7 - 6\times10^8 \: M_\sun$, with the
median dust mass being $1.6\times10^8 \: M_\sun$.  This is a factor of nearly 20
larger than the median dust mass inferred from the measurements of $A_V$.  This
might be expected, given that many of the dust masses based on $A_V$ are lower
limits, while the dust masses based on the 24$\mu$m emission may be
overestimates if $T_{\rm d} > 75$~K.  On the other hand, this difference may be
suggesting that the dust causing the average UV extinction of the extended
galaxy component is not the same dust that is causing the thermal emission.

We use the SSP templates to estimate the stellar mass in each DOG.  This is
computed by reddening each SSP template to match the observed color of the DOG
at the appropriate redshift.  We then scale the redshifted, reddened template
to match the observed H-band photometry.  
Since the \citet{2003MNRAS.344.1000B} models are normalized to a stellar mass
of 1~$M_\sun$, this scaling factor represents the stellar mass of the DOG.  

In Figure~\ref{fig:mstar}, we show the stellar mass of each DOG as a function
of age as well as the distribution of stellar masses assuming a 100~Myr SSP
model.  As stellar populations age, their colors naturally redden, and so less
extinction is needed to reproduce the observed colors of the DOGs.  For most
DOGs, ages greater than $\sim$300~Myr require $A_V$ values less than zero and
are unphysical.  Meanwhile, at younger ages, lower mass-to-light ratios are
balanced by the need for greater extinction to match the observed colors.  As a
result, the inferred stellar masses are relatively constant to within a factor
of a few for ages less than 300~Myr.  We note that these mass estimates are
lower limits because (1) the amount of extinction is a lower limit, especially
when there is no detection in the $V$- or $I$-band image and (2) our extinction
estimate does not take into account grey extinction.  For an age of 100~Myr,
the stellar masses range from $2\times10^8-1\times10^{12} \: M_\sun$, with the
median mass being $3.3\times10^{10} \: M_\sun$.

\begin{figure*}[!tbp]
\epsscale{1.20}
\plotone{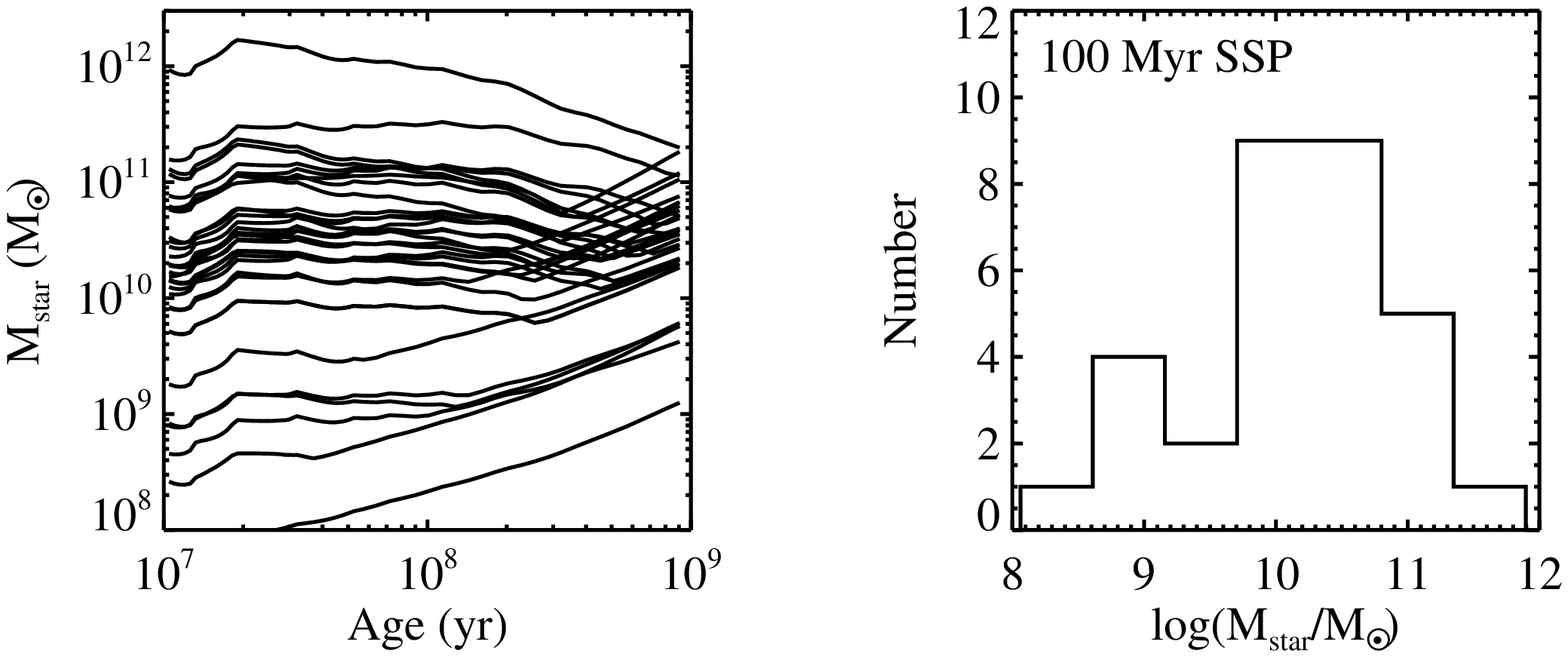}

\caption{ {\it Left}: Stellar mass as a function of SSP age for the DOGs.  {\it
Right}: Distribution of stellar masses at an age of 100~Myr.}

\label{fig:mstar}

\end{figure*}

We use our dust mass estimates inferred from the 24$\mu$m flux densities and a
gas-to-dust mass ratio of 120 \citep{2008arXiv0806.3002W} to obtain an upper
limit on the gas masses.  This leads to gas masses of 1-7$\times10^{10} \:
M_\sun$, with a median gas mass of $2\times10^{10}$.  If we assume a closed-box
model and an exponential star-formation history, then we can write $M_{\rm gas}
+ M_{\rm dust} = M_{\rm tot} {\rm exp}(-t/t_0)$, where $M_{\rm tot}$ is the sum
of the gas, dust, and stellar masses.  The stellar mass is then given by $M_{\rm
star} = M_{\rm tot} (1 - {\rm exp}(-t/t_0))$, which implies 

\begin{equation}
\frac{t}{t_0} = {\rm ln}\frac{M_{\rm tot}} {M_{\rm gas} + M_{\rm dust}}.
\label{eq:tt0}
\end{equation}

This quantity represents the fractional lifetime (in units of the scale time
for our exponential star-formation rate assumption) of each DOG.  Larger values
of $t/t_0$ indicate more evolved systems, as more of the gas has been converted
to stars.  Our values of $t/t_0$ represent lower limits on the actual values
because our stellar masses are underestimated and our gas masses are based on
our 24$\mu$m flux densities, which likely overestimates the true mass of gas.
We find $t/t_0$ lower limits ranging from 0.02 to 3.3, with a median lower
limit of 0.9.  This result implies that in half of the DOGs in our sample, at
least 90\% of one exponential timescale's worth of star-formation has occurred.
The ``oldest'' DOG in our sample has gone through more than three exponential
timescales of evolution.  However, we caution that the ``youngest'' DOGs from
this line of analysis are uniformly associated with sources where the dust
extinction is most likely underestimated, thereby causing an additional
underestimate in the stellar mass and the associated $t/t_0$ value.  In
Table~\ref{tab:masses}, we present the dust and stellar masses derived in this
section, as well as our measure of the lower limit on the fractional lifetime,
$t/t_0$. 

\begin{deluxetable*}{lccc}
\tabletypesize{\scriptsize} 
\tablecolumns{4}
\tablewidth{240pt}
\tablecaption{DOG Mass Estimates}
\tablehead{
\colhead{} & \colhead{$M_{\rm dust}$\tablenotemark{a}} &  
\colhead{$M_{\rm star}$\tablenotemark{b}} & \colhead{} \\
\colhead{} & \colhead{(10$^7 \: M_\sun$)} & 
\colhead{(10$^{10} \: M_\sun$)} & \colhead{$t/t_0$\tablenotemark{c}}
}
\startdata
SST24 J142538.2+351855 & $0.8-11$ &  0.8 &  0.5 \\
SST24 J142622.0+345249 & $0.5-13$ &  0.4 &  0.2 \\
SST24 J142626.4+344731 & $2.0-15$ &  2.9 &  1.0 \\
SST24 J142644.3+333051 & $0.3-32$ &  11. &   1.4 \\
SST24 J142645.7+351901 & $5.2-13$ &  5.1 &   1.5 \\
SST24 J142648.9+332927 & $0.7-24$ &  1.3 &  0.4 \\
SST24 J142653.2+330220 & $2.9- 8$ &  3.8 &   1.6 \\
SST24 J142804.1+332135 & $2.8-12$ & 0.02 &  0.02 \\
SST24 J142924.8+353320 & $0.2-16$ & 0.08 &  0.04 \\
SST24 J142958.3+322615 & $0.5-19$ &  2.0 &  0.6 \\
SST24 J143001.9+334538 & $0.1-60$ &  0.1 &  0.02 \\
SST24 J143025.7+342957 & $0.3-37$ &  2.8 &  0.5 \\
SST24 J143102.2+325152 & --- & ---       &  ---  \\
SST24 J143109.7+342802 & $7.7-13$ &  2.2 &  0.9 \\
SST24 J143135.2+325456 & $4.9-29$ &  4.5 &  0.8 \\
SST24 J143225.3+334716 & $1.7-13$ &  9.6 &   2.0 \\
SST24 J143242.5+342232 & $4.7-11$ &  13. &   2.4 \\
SST24 J143251.8+333536 & $3.5- 9$ &  13. &   2.6 \\
SST24 J143312.7+342011 & $3.6-23$ &  4.9 &   1.0 \\
SST24 J143325.8+333736 & $3.8-17$ &  32. &   2.8 \\
SST24 J143358.0+332607 & $1.1-15$ &  6.6 &   1.5 \\
SST24 J143447.7+330230 & $0.9-18$ &  2.4 &  0.7 \\
SST24 J143504.1+354743 & $5.1-16$ &  5.4 &   1.3 \\
SST24 J143508.4+334739 & $0.4-31$ &  1.4 &  0.3 \\
SST24 J143520.7+340418 & $0.1-16$ &  0.1 &  0.05 \\
SST24 J143523.9+330706 & $0.5-16$ &  3.3 &   1.0 \\
SST24 J143539.3+334159 & $1.1-43$ &  2.0 &  0.3 \\
SST24 J143545.1+342831 & $2.5-30$ &  95. &   3.3 \\
SST24 J143644.2+350627 & $1.0-22$ &  3.7 &  0.9 \\
SST24 J143725.1+341502 & $2.4-22$ &  13. &   1.8 \\
SST24 J143808.3+341016 & $1.1-26$ &  12. &   1.6 \\
\tablenotetext{a}{Mass range reflects estimates based on $A_V$ and 24$\mu$m
flux density}
\tablenotetext{b}{Stellar mass estimates represent lower limits on true stellar
mass}
\tablenotetext{c}{$t/t_0$ estimates are lower limits based on 24$\mu$m dust masses}
\enddata
\label{tab:masses}
\end{deluxetable*}


\subsection{Comparison to other high redshift galaxy populations}
\label{sec:comp}

It is important to understand how DOGs are related to other populations of
high-$z$ galaxies that have been studied in the literature.  Here we compare
the morphological properties of the DOGs with some of these high-redshift
galaxy populations and find that DOG morphologies are distinct from the bulk of
LBGs and quiescent high-$z$ galaxies, but are similar to SMGs as well as active
DRGs and the extreme subset of faint, diffuse LBGs.  

\subsubsection{Sub-mm Galaxies}\label{sec:smgs} SMGs are a particularly
interesting population of galaxies to compare with DOGs.  First identified by
blind sub-mm surveys with the Submillimetre Common User Bolometer Array
\citep[SCUBA][]{1999MNRAS.303..659H}, SMGs may represent an important,
short-lived, and very active phase in the evolution of the most massive
galaxies.  Their redshift distribution, number density, and clustering
properties are similar to DOGs
\citep{2005ApJ...622..772C,2008ApJ...677..943D,2004ApJ...611..725B,2008arXiv0810.0528B}.
However, a sample of DOGs detected at 70$\mu$m or 160$\mu$m by {\it
Spitzer}/MIPS tend to show warmer colors (i.e., smaller 70/24$\mu$m or
160/24$\mu$m flux density ratios) compared to SMGs (Tyler et al., submitted).
One speculative scenario that may serve as a possible explanation for this
behavior is that these two galaxy populations are linked in an evolutionary
sense: SMGs represent a cold dust, star-formation dominated stage in the
formation of massive galaxies that may precede the DOG phase, when the feedback
from the growth of a central black hole has heated the surrounding gas and
dust, thereby quenching star formation and shifting the peak of the SED to
shorter wavelengths.  

If this scenario is correct, then we expect to see major mergers dominate the
morphologies of SMGs, while DOGs should show more relaxed morphologies typical
of the final merger stage before the remnant.  \citet{2003ApJ...596L...5C}
analyzed Space Telescope Imaging Spectrograph (STIS) rest-frame UV data of a
sample of 11 SMGs at $z \sim 2-3$ using the CAS system, and found evidence
suggesting a major merger fraction of between 40\% and 80\%.  Although we do
not have the S/N in our images to measure $A$ reliably (and thereby determine a
major merger fraction in a similar manner), the low $G$ and high $M_{20}$
values we have measured imply diffuse, irregular systems where the light is
spread into multiple components rather than 2 separate components.  If the DOGs
were predominantly major mergers, we would expect our sample to have higher $G$
values.  Instead, the low $G$ values we find suggest that we may be looking at
the dusty remnant of a major merger, where there are many highly obscured
components near each other.  However, we caution that dust can have a strong
effect on the measured $G$ value in the rest-UV, such that even major mergers
might yield lower $G$ values.

The rest-UV morphologies of SMGs have also been analyzed in the GOODS-N field,
where a SCUBA super-map exists and has been used to identify robust sub-mm
detections \citep{2003MNRAS.344..385B}.  A sample of 12 sources in the redshift
range 1.7-4.0 (comparable to the DOGs) were studied by
\citet{2005MNRAS.358..149P}, who computed concentration and asymmetry values,
finding $C$ to be in the range 2-3.3 and $A$ to be dominated by noise, with the
exception of two objects (one is very compact and the other is clearly
asymmetric).  The comments associated with many of these sources are ``faint''
and ``diffuse'', suggesting that there is large-scale dust obscuration in these
systems.  This is qualitatively similar to what is seen in many of the DOGs,
suggesting that there is some overlap between the two samples.  

SMGs have not yet been characterized in terms of $G$ or $M_{20}$, so direct
comparisons based on these quantities are not possible at this time.  However,
we can compare the sizes of these systems directly via the Petrosian radius.
In the rest-UV, the DOGs range in size from $R_p \sim 0.5\arcsec$ to
1.5$\arcsec$, while SMGs range in size from 0.5 - 2.5$\arcsec$.  Indeed, a
two-sided KS test reveals that there is only a 5\% chance that they are drawn
from the same parent distribution.  This suggests that, while there are
similarities between SMGs and DOGs, SMGs tend to be slightly larger than DOGs.
This is consistent with the major merger hypothesis in which DOGs are in a more
evolved state where dynamical friction has caused individual components to fall
towards the center of mass. However, we caution that this result in itself does
not provide evidence for DOGs originating from major mergers.  Objects moving
at $\sim$100~km~s$^{-1}$ will traverse 8~kpc in $\lesssim$100~Myr.  Simulations
of major mergers predict a phase of intense star formation and central black
hole growth that lasts of order this timescale, indicating that the size
differences are at least consistent with the scenario outlined above
\citep{2007arXiv0706.1243H}.  

Finally, the ratio of the stellar mass to the gas mass holds potential for
comparing the evolutionary states of SMGs and DOGs.  The ideal comparison study
would include statistically significant samples of both populations of galaxies
for a range of observed properties such as 24$\mu$m emission, bolometric
luminosity, space density, etc.  Unfortunately, such samples do not currently
exist --- mainly due to a combination of the limitations of current
instrumentation and the fact that these are relatively recently discovered
populations of galaxies.  While CO linewidths and emission strengths have found
gas mass estimates for a handful of SMGs, no such measurements have been
published for DOGs.  The gas mass estimates for SMGs are typically
$\sim$5$\times10^{10} \: M_\sun$
\citep{2005MNRAS.359.1165G,2006ApJ...640..228T,2008ApJ...680..246T}.  A number
of efforts have been directed at determining the stellar masses of SMGs using
SED-fitting algorithms.  Average stellar mass values are in the range
3-6$\times10^{11} \: M_\sun$ \citep{2005ApJ...635..853B,2008MNRAS.386.1107D}.
However, when we employ our method of determining the stellar mass (in this
case using the $R - K$ color to determine the optimal $A_V$ value for a given
SSP and age) using the photometry presented in those papers, we find average
stellar masses of $\sim$7$\times10^{10} \: M_\sun$.  These estimates increase
by a factor of $\approx$2 if instead of a Chabrier IMF we use a Salpeter IMF
(as was done by the previous authors for SMGs).  However, this still leaves us
a factor of $\approx$2 short of the mass estimates provided in the papers
described above.  In order to compare DOG stellar masses with SMGs
consistently, we adopt the lower stellar mass values that we derive for SMGs.
In this case, the median $t/t_0$ values for SMGs becomes $\approx$1.1, which is
much closer to the median lower limit value of 0.9 found for the DOGs in
section~\ref{sec:masses}.

The large uncertainty inherent in the process of estimating dust and gas masses
based on 24$\mu$m photometry or rest-frame optical $A_V$ measurements currently
prevents a strong conclusion being made regarding the evolutionary status using
this line of analysis.  However, the morphological evidence is suggestive ---
although not conclusive --- of an evolutionary link between the two populations
with SMGs serving as the less evolved precursor to the DOG phase.

\subsubsection{Star-forming Galaxies}\label{sec:starform}  A number of
selection criteria have been used to identify normal star-forming galaxies at
high redshift.  Two of these are the LBG dropout
\citep{1996MNRAS.283.1388M,1996ApJ...462L..17S} and $BzK$
\citep{2004ApJ...617..746D} techniques.  A direct comparison to our work can be
made with a sample of LBGs and emission line galaxies in the GOODS-N field
studied by \citet{2006ApJ...636..592L}.  These authors compared $G$, $M_{20}$,
and $C$ values between their sample of 82 $z \sim 4$ LBGs and 55 $z \sim 1.5$
emission line galaxies.  In the LBG sample, they found a major-merger fraction
of $\sim 10-25\%$ (defined by $M_{20} \geq -1.1$) and a bulge-dominated
fraction of $\sim 30\%$ ($G \geq 0.55$,~$M_{20} < -1.6$).  The remainder of the
LBGs had $G$ and $M_{20}$ values larger than what is typical for normal
galaxies, suggesting active star-formation or a recent merger event.  The
low-$z$ emission-line sample showed a similar major merger fraction but fewer
bulge-dominated systems.  It is remarkable then, that so few of the DOGs have
$G$ and $M_{20}$ values typical of bulge-dominated systems, even in the
rest-optical, despite their luminosity.  Furthermore, four out of six DOGs with
measureable morphologies in the rest-UV have high $M_{20}$ and low $G$ values
that are typical of dusty, irregular systems.  This may be an indication of kpc
scale dust obscuration, which can bias the $G$ and $M_{20}$ values away from
the bulge-dominated regime.

A morphological study of LBGs in GOODS-N by \citet{2006ApJ...652..963R} found
axial ratios skewed towards lower values for galaxies at $z > 3$, suggesting
high-$z$ LBGs are dominated by edge-on morphologies.  In contrast, only one DOG
has an axial ratio less than 0.35, indicating that if these sources are disk
galaxies, then some selection mechanism must be in place that favors observing
DOGs in face-on orientations.  Meanwhile, results from numerical simulations of
galaxy mergers indicate that remnants end up with axial ratios between 0.5 and
1.0, depending on the viewing angle \citep{2006ApJ...646L...9N}.  21 DOGs
satisfy this axial ratio criterion, but the median value in our dataset is
$\approx$0.5.  This suggests that either DOGs represent a phase prior to the
final remnant stage or they are formed by some other process.

Recently, \citet{2007ApJ...669..929L} have used GOODS data to analyze
morphologies of 216~LBGs and compare them with other high-$z$ galaxy
populations.  They found significant overlap between the LBGs and $BzK$s,
indicating that the optical and NIR selection criteria are identifying similar
galaxies.  While these authors performed a non-parametric morphological
analysis, direct comparison between our work and theirs is difficult because
(a) they assign pixels to each galaxy based on an isophotal surface brightness
criterion rather than the elliptical Petrosian radius as we have done and (b)
they create their own parameter to describe the multiplicity ($\Psi$) of each
galaxy, rather than using $M_{20}$.  Nevertheless, there are some apparent
differences between the DOGs and LBGs from the \citet{2007ApJ...669..929L}
study.  While the LBGs span the full range of $G$ values in the rest-UV, the
DOGs tend to be low $G$ objects.  Furthermore, though LBGs span a wide range in
$G$, they are preferentially found to have low $\Psi$ values, implying a
small number of distinct components.  On the other hand, the DOGs have high
$M_{20}$ values, suggesting multi-component structure is commonplace.
\citet{2007ApJ...669..929L} note a correlation in their plot of $G$ as a
function of $\Psi$, in the sense that objects with many components (large
$\Psi$) tend to be fainter and more nebulous (low $G$).  DOGs resemble this
extreme subset of faint diffuse LBGs, but appear highly morphologically
distinct from the vast majority of the LBG population.

\subsubsection{Passively Evolving Galaxies}\label{sec:passive}  As mentioned
above, the $BzK$ method can be used to identify high-$z$ passively evolving
galaxies.  This photometric color cut has been used in the Hubble Ultra Deep
Field (UDF) to generate a sample of seven luminous early-type galaxies at $z
=1.39 - 2.47$ \citep{2005ApJ...626..680D}.  These authors studied the $i$ and
$z$ band morphologies of all seven objects with both parametric (S\'ersic
profile fitting) and non-parameteric (concentration and asymmetry) methods.
They found fairly large S\'{e}rsic indices ($n\sim3$) and small effective radii
($r_{\rm eff} \lesssim 1 \:$kpc), typical of E/S0 galaxies.  In contrast, the
best-fit S\'{e}rsic profile for DOGs has smaller $n$ values more typical of
exponential disks (median $n = 0.9$) and larger effective radii ($R_{\rm eff}
\sim 1-6 \:$kpc).  Moreover, the passive $BzK$ galaxies have $C > 2.6$ and $A <
0.2$, consistent with early-type systems.  In the rest-optical, DOGs tend to
show lower $C$ values (S/N is not sufficient to measure $A$), consistent with
an exponential profile.  Along with the low $G$ and high $M_{20}$ values that
are measured for the DOGs, these morphology results suggest that DOGs and
passively evolving high-$z$ galaxies are distinct populations, either because
they represent different stages of evolution or because they have different
formation mechanisms.

\subsubsection{Distant Red Galaxies}\label{sec:drg}  Another population of
high-$z$ galaxies is the so-called Distant Red Galaxies (DRGs).  Identified via
deep NIR imaging, these objects were first postulated to be the reddened
descendents of LBGs \citep{2003ApJ...587L..79F}.  Subsequent studies of DRGs in
the Extended Groth Strip (EGS) show a wide variety of shapes, with 57\%
appearing visually as elliptical/compact, 7\% as edge-on disks, and the
remainder as peculiar/irregular galaxies \citep{2007ApJ...660L..55C}.  The low
redshift DRGs ($z < 1.4$) have $CAS$ values typical of nearby normal galaxies.
The higher $z$ DRGs visually classified as elliptical/compact have higher $C$
values, similar to what is seen locally in massive ellipticals and in the $BzK$
samples.  Meanwhile, \citet{2007ApJ...669..929L} examined DRGs in the GOODS-N
field that did not overlap with the BX or BM LBG color criteria and found that
this population of galaxies was substantially fainter and more diffuse than
either the star-forming $BzK$s or the LBGs.  They note that this is the
behavior one expects from dusty, IR-bright galaxies.  The faint, diffuse nature
of these objects is reminiscent of the DOGs, and it is possible that there is
significant overlap between these two populations.

Previous work using rest-frame UV-NIR SEDs has separated actively star-forming
DRGs (sDRGs) from quiescent ones (qDRGs) \citep{2007ApJ...656...66Z,
2007ApJ...671..285T}.  Examination of the morphological differences between
these two populations has revealed a correlation between size and star formation
activity, in the sense that qDRGs are all very small ($R_{\rm eff} \lesssim 1
\:$kpc), while sDRGs span a larger range in size ($R_{\rm eff} \sim 1 - 10
\:$kpc).  As is shown in the right panel of Figure~\ref{fig:rpetmag}, DOGs
appear very similar to sDRGs in terms of their sizes.  This is consistent with
the qualitative similarity between sDRGs and DOGs described in the preceding
paragraph and suggests there is extensive overlap between these two populations.

\subsection{Implications for the Evolution of the Most Massive
Galaxies}\label{sec:impl}  

In the local universe, there has been evidence for some time that warm
dust-dominated ULIRGs may represent a transition stage between cold ULIRGs and
optically luminous quasars \citep{1988ApJ...328L..35S}.  If this scenario holds
at high redshift, then there is a natural explanation for the observations
based on the selection criteria alone: objects selected at long wavelengths
(i.e., SMGs) are preferentially cold-dust dominated systems and represent the
`cold ULIRG' phase, whereas objects selected at 24$\mu$m (i.e., DOGs) are
dominated by warmer dust and represent the transition phase en route to the
optically luminous quasar.  As time progresses and the quasar fades in
luminosity, the compact, quiescent, elliptical galaxy remnant becomes visible
(i.e., quiescent $BzK$s and DRGs).  

Again referring to the local universe for guidance, if the triggering mechanism
for this activity is a major merger \citep{1988ApJ...325...74S}, then we should
expect to see a trend in relaxation and size, where the initial stage shows the
largest sizes and least relaxation and the end product is a relaxed, compact
system.  This picture is apparently consistent with our data, as DOGs tend to
be smaller than SMGs, but larger than quiescent DRGs or $BzK$ galaxies.
Furthermore, SMGs frequently exhibit signs of major merger activity, whereas
passively evolving systems at high-$z$ are very compact with large S\'{e}rsic
indices.  DOGs appear to be intermediate stage objects that typically do not
show signs of major mergers, but nonetheless have morphologies indicating they
are more dynamically relaxed than SMGs but less so than the quiescent systems.  

It is important to emphasize that while our morphological results are
consistent with the hypothesis that DOGs act as a transition phase in the
process of creating a massive galaxy via a major merger, the morphological
information currently available is not sufficient to exclude the possibility
that DOGs are created by some other process such as minor merging (for example,
minor mergers have the potential to increase size temporarily), or are simply
dusty galaxies hosting a powerful, obscured AGN.

Our analysis of the stellar, dust, and gas masses of DOGs currently does not
provide compelling evidence to place them within an evolutionary scheme with
respect to other massive proto-galaxy candidates such as SMGs.  Additional data
are needed before conclusive statements can be made based on mass estimates
such as these.

\section{Conclusions} \label{sec:conclusions}

We have analyzed the morphologies of 31 Dust Obscured Galaxies (DOGs) at $z
\approx 2$ from the Bo\"{o}tes field using data from {\it HST} ACS/WFPC2 and
NICMOS.  Our findings are summarized below.

\begin{enumerate}
		


\item Although these sources were selected to have mid-IR signatures of AGN,
we detect spatially resolved emission at rest-frame UV and/or rest-frame
optical wavelengths for all but one of the 31 targets.

\item Using a three component model in GALFIT (sky + PSF + S\'ersic profile),
we measure significant unresolved components in 28 out of 31 DOGs in the
rest-optical, and the median point-source fraction is 0.13.  Only 10 DOGs have
measureable unresolved components in the rest-UV.    
	
\item The median S\'ersic index is 0.9, indicating that disk-like profiles are
preferred to bulge-like ones.  On the other hand, very few DOG extended
components have small axial ratios, indicating that if DOGs are predominantly a
population of normal, disk-like galaxies (with an obscured AGN producing the
24$\mu$m flux), then some selection mechanism(s) must be in place that favors
face-on rather than edge-on orientations.

\item DOGs in our sample have effective radii of 1-5~kpc, which places them
between SMGs and quiescent DRGs or $BzK$ galaxies.  If DOGs are formed by a
major merger, this trend in sizes is consistent with them acting as a
transition stage in the evolution of massive galaxies.  If DOG activity is
triggered by some other process, such as a minor merger or a dusty AGN in a
normal galaxy, then interpretation of this size trend is not as clear.

\item In the rest-optical, DOGs have lower $G$ values than local ULIRGs (median
values of 0.49 and 0.59, respectively).  This might be expected if DOGs
represent a subsequent stage in the merging process (just before coalescence),
but might also be expected if the galaxies are not disturbed by a major merger.  

\item Simple stellar population modeling reveals that old ($>$300~Myr)
single-burst stellar populations are redder than most DOGs and thus ruled out.
If 100~Myr old SSPs are appropriate, then DOGs require substantial amounts of
extinction to produce the observed red colors, with $A_V=0.2-3$.  This provides
a lower bound on the median dust mass of $10^7 \: M_\sun$.  An upper bound is
obtained by extrapolating the 24$\mu$m flux density to 850$\mu$m and is found
to have a median value of $1.5 \times 10^8 \: M_\sun$.  We find a median
stellar mass lower limit of $3 \times 10^{10} \: M_\sun$ which is relatively
insensitive to age to within a factor of a few.

\end{enumerate}

This work is based in part on observations made with the {\it Spitzer Space
Telescope}, which is operated by the Jet Propulsion Laboratory, California
Institute of Technology under NASA contract 1407. We are grateful to the expert
assistance of the staff Kitt Peak National Observatory where the Bo\"{o}tes
field observations of the NDWFS were obtained. The authors thank NOAO for
supporting the NOAO Deep Wide-Field Survey. In particular, we thank Jenna
Claver, Lindsey Davis, Alyson Ford, Emma Hogan, Tod Lauer, Lissa Miller, Erin
Ryan, Glenn Tiede and Frank Valdes for their able assistance with the NDWFS
data.  We also thank the staff of the W.~M.~Keck Observatory, where some of the
galaxy redshifts were obtained.

RSB gratefully acknowledges financial assistance from HST grant GO10890,
without which this research would not have been possible.  Support for Program
number HST-GO10890 was provided by NASA through a grant from the Space
Telescope Science Institute, which  is operated by the Association of
Universities for Research in  Astronomy, Incorporated, under NASA contract
NAS5-26555.  The research activities of AD and BTJ are supported by NOAO, which
is operated by the Association of Universities for Research in Astronomy (AURA)
under a cooperative agreement with the National Science Foundation.  Support
for E. Le Floc'h was provided by NASA through the Spitzer Space Telescope
Fellowship Program.

\end{document}